\magnification=1200
\vsize=8.5truein
\hsize=6truein
\baselineskip=20pt
\centerline{\bf The structure and phase transitions in} 
\centerline{\bf polymer blends, diblock copolymers and liquid crystalline 
polymers:} 
\centerline{\bf the Landau-Ginzburg approach}
\vskip 20pt
\centerline{by}
\vskip 20pt
\centerline{Robert Ho{\l}yst and T.A.Vilgis$\ddagger$}
\vskip 20pt
\centerline {Institute of Physical Chemistry PAS and College of Science,
 Dept. III,} 
\centerline {Kasprzaka 44/52, 
01224 Warsaw, Poland}
\centerline{$\ddagger$ Max-Planck-Institut f{\"u}r  Polymerforschung,
P.O. Box 3148, D-55021 Mainz, F.R.G. }
\vskip 60pt
\centerline{\bf{Abstract}}
The polymer systems are discussed in the framework of
the Landau-Ginzburg model. The model is derived from the 
mesoscopic Edwards hamiltonian via the conditional
partition function. We discuss flexible, semiflexible and
rigid polymers. The following systems are studied:
polymer blends, flexible diblock and multi-block copolymer melts,
random copolymer melts, ring polymers,
rigid-flexible diblock copolymer melts,
mixtures of copolymers and homopolymers and mixtures of 
liquid crystalline polymers.
Three methods are used to study the systems: mean-field model,
self consistent one-loop approximation and self consistent
field theory. The following problems are studied and discussed:
the phase diagrams, scattering intensities and correlation
functions, single chain statistics and behavior of single chain
close to critical points, fluctuations induced 
shift of phase boundaries. 
In particular we shall discuss shrinking of the polymer chains 
close to the critical point in polymer blends, size of the
Ginzburg region in polymer blends and shift of the critical 
temperature. 
In the rigid flexible diblock copolymers we shall
discuss the density nematic order parameter correlation function.
The correlation functions in this system are found to oscillate with
the characteristic period equal to the length of the rigid part of the diblock
copolymer. The density and nematic order parameter measured along the
given direction are anticorrelated. 
In the flexible  diblock copolymer system we shall discuss
various phases including the double diamond and gyroid
structures. The single chain statistics in the 
disordered phase of flexible diblock
copolymer system is shown to deviate from the Gaussian statistics due
to fluctuations. In the one loop approximation one shows that
the diblock copolymer chain is stretched in the point where two incompatible 
blocks meet but also that each block shrinks close to the microphase
separation transition. The stretching outweights shrinking and the
net result is the increase of the radius of gyration about the
Gaussian value.
Certain properties of 
the homopolymer, copolymer system are discussed.
The diblock copolymers solubilizes
two incompatible homopolymers by forming a monolayer
interfaces between them. The interface has the positive saddle splay
modulus which means that the interfaces in the disordered phase 
should be characterized by the negative Gaussian curvature.
We also show that in such a mixture the Lifshitz tricritical point 
is encountered. The properties of this
unusual point are presented. The Lifshitz, equimaxima and disorder
line are shown to provide a useful tool for studying local
ordering in polymer mixtures.
In the liquid crystalline mixtures 
the isotropic nematic phase transition
is discussed. We concentrate on
static, equilibrium properties of the polymer systems. 
\vfill\eject
\centerline{\bf I. Introduction}  

The introduction of the
order parameter in physics was a milestone in the description of
phase transitions. The same Landau-Ginzburg 
approach has been applied to such diverse 
phenomena as transitions in liquid crystals and
superconductivity, liquid-solid or solid-solid phase transitions and 
transitions in the superfluid helium, ferromagnetic paramagnetic phase
transition and phase transitions in polymer systems. 
The purpose of this review is  the application of the
Landau-Ginzburg model to polymer systems.  
Despite the obvious limitations of this approach (valid for weakly first
order and continuous phase transitions) it provides plenty of information
about the polymer systems such as: polymer blends, diblock copolymers and
liquid crystalline polymers.  In polymer systems the
parameters characterizing the Landau-Ginzburg free energy can be
directly related to the mesoscopic parameters of the polymer systems.
The mesoscopic parameters characterizing for example the interactions
are, on the other hand,  related to 
the microscopic interaction parameters via the
integrals of the direct corelation functions. Therefore  the results 
obtained in
the Landau-Ginzburg model shed light on the 
relation between the mesoscopic and/or microscopic
parameters of the system and its
behaviour.

The simplest 
polymer mixture is the A homopolymer, B homopolymer blend. The
very fruitful approach to the study of the system
near its consolute (critical) point 
is based on the mean field Flory-Huggins free energy$^{1)}$
and the de Gennes
random phase approximation (RPA)$^{2)}$ for the scattering
intensity. Both can be stated in terms of the 
Landau-Ginzburg model
as we show in Section III.
RPA and Flory-Huggins approach are two compatible mean-field
models. They are named mean-field since  in both models only
the most probable configurations of the polymer
chain are taken into account and, more importantly,
the critical long-wavelength fluctuations
are neglected.  
It was de Gennes who pointed out$^{3)}$ that 
the mean-field theory is rather good for high molecular mass polymer
mixtures, in contrast to the low molecular 
mass mixtures, 
for which the mean-field theory breaks down close to the critical
point. In fact, in the limit of $N_A,N_B\to\infty$, the mean-field
theory is exact$^{3-5)}$.
The mean-field approach is quantitatively
correct if 
the correlation length $\xi\sim (T/T_c-1)^{-1}$ is comparable or smaller
than the typical length scale in the system i.e. the radius of gyration
$\sim\sqrt{N}$. 
Comparing two length scales, for
large $N$, we find that the
mean-field theory breaks down very close to the
critical point, i.e. for
 $$\vert T-T_c\vert/T_c\sim 1/N.\eqno(1.1)$$
This is the Ginzburg criterion. The region around the critical point
where the mean field theory breaks down is called the Ginzburg region. 
The shift of the critical temperature, $T_c$, with respect to $T_c^{mf}$  
(calculated in the Flory Huggins model)   
is 
$$\vert T_c-T_c^{mf}\vert/T_c^{mf}\sim
1/\sqrt{N}\eqno(1.2).$$ 
The shift, induced by fluctuations, is 
obtained in the one-loop approximation$^{6,7)}$.  It
 is larger than the Ginzburg region itself. 
The  derivation of Eq(1.2) is based on
the study of the scattering intensity, $I(q)$, where $q$ is the scattering
wavevector. For $T\to T_c$ and $q\to 0$, we expect 
$I^{-1}(0)\sim (T/T_c-1)^{-\gamma}$, 
where $\gamma =1$ in the mean-field theory$^{6,7)}$.
The breakdown of this 
relation is usually used in experiment 
to estimate the Ginzburg region$^{7,8-13)}$ and  therefore, we  also
base the calculations of the Ginzburg region
on the scattering intensity$^{14,15)}$. 
Sufficiently close to $T_c$, fluctuations 
become important
and $\gamma$ changes from 1 to its universal value of 1.26. This change
can be extracted if one plots $\log I(0)$ versus $\log \vert T/T_c-1\vert$. 

Different polymers can be combined into single material in many ways and
polymer blends find widely spread application. Many of the potential 
application depends on the morphology of the system which in turn depends
on the single chain behaviour. Here we shall discuss the 
behavior of the single chain in the polymer blend$^{16-19)}$ as well as in the
diblock copolymer system$^{20,21)}$. 
The chains shrink considerably at the approach to the critical point in homopolymer
blends. Intuitively we can expect that when the correlation length becomes
larger than the radius of gyration the chains demix on this scale$^{18)}$.
On the other hand the chains stretch in diblock copolymer system$^{20,21)}$.
 
An A-B diblock copolymer is a polymer consisting of a sequence of A-type
monomers chemically joined to a sequence of B-type monomers. Even a small
amount of incompatibility (difference in interactions) between monomers A and
monomers B can induce phase transitions in a mixture of homopolymers. However
A-homopolymer and B-homopolymer are chemically joined in a diblock, so that a
system of diblocks cannot undergo a macroscopic phase separation. Instead a
number of order-disorder phase transitions take place in the system between
the isotropic phase and spatially ordered phases in which A-rich and B-rich
domains, of the size of a diblock copolymer, are periodically 
arranged in lamellar, hexagonal, bcc, lamellar-catenoid,
double diamond (Fig.1) or gyroid structures (Fig.2). 
A Landau-Ginzburg model of the phase transitions in diblock 
copolymer system was first 
formulated by Leibler$^{22)}$ and later refined 
by Fredrickson et al$^{20,23)}$.
Matsen and Schick$^{24,25)}$ applied the self consistent field theory$^{26,27)}$
 to the
phase transitions in this system in the weak segregation limit$^{28)}$, where
the domain boundaries between the 
A rich and B rich domains are wide. The opposite case is the strong segregation
regime with very sharp and narrow domain boundaries; here we consider 
only the former
case.  The theoretical studies of the system were greatly stimulated by
experiments. The cylindrical and spherical and lamellar phases
have been known for long time$^{29)}$, but only
recently the novel bicontinuous structures have been discovered 
in the diblock copolymer systems$^{30-32)}$ (Fig.1).
In the polystyrene-polyisoprene system (Fig.3) the bicontinuous double
diamond structure consists of two channels each of diamond symmetry
separated by the surface (Fig.1), which as has been argued by Thomas
et al$^{32)}$ is of constant mean curvature$^{33)}$.

The richness of the phase diagram and behavior of the polymer system is
enhanced if we mix the diblock copolymers and homopolymers$^{34-37)}$. 
Ternary mixtures of A,B homopolymers and A-B diblock copolymer$^{34-36)}$ 
are similar in some respects to 
mixtures of oil, water and amphiphile$^{38)}$.
The copolymer acts as a surfactant for the A,B homopolymer blend, decreasing
the surface tension between A-rich and B-rich phases as it accumulates
on the interface between these two phases and reduces the number of
energetically unfavorable contacts between the A-homopolymers and 
B-homopolymers. In spite of this similarity, the comopolymer is not generally 
as efficient a solubilizer of homopolymers as a good amphiphile is of oil and 
water$^{35)}$. The analysis of the disordered system is based on the study of
disorder$^{39)}$, Lifshitz and equimaxima lines.
The disorder line 
of the system is the locus of points at which all 
correlation functions in 
general no longer 
decay monotonically, but contain an exponentially damped {\it oscillatory }
component. This oscillation reflects the tendency of the copolymer to order 
the A (B) monomers of the system.  The Lifshitz line is the locus of 
points at which the peak in the structure function just begins to move 
off of zero 
wavevector. It therefore indicates the point at which oscillatory components 
begin to {\it dominate} the {\it particular} correlation function, in contrast 
to the disorder line which indicates the point at which oscillatory components 
{\it appear} in {\it all} correlation functions. The Lifshitz 
line of the structure function of all A monomers is quite close to the 
disorder line, indicating that once the tendency to order the system appears, 
it quickly dominates the response of this function. In contrast, 
the Lifshitz line associated 
with the structure function of those monomers located only on the homopolymers
is, in a large part of the phase diagram, far from the disorder line. 
These results taken together 
indicates that the copolymer is rather ineffective in organizing the 
homopolymers, and much more efficient in organizing itself, an 
organization expressed most strongly in the lyotropic phases which easily 
exist in the complete absence of homopolymer.                                  
Beyond the disorder line 
the interface between the A-rich homopolymer B-rich homopolymer phase 
is not wetted by the disordered phase$^{36)}$ so it consists of
of a thin copolymer layer. 
In the disordered phase we expect that A-rich and B-rich domains are 
also separated
by the thin copolymer layer. 
Assuming that the structure of the interface 
is the same as the internal interfaces in the disordered phase 
Matsen and Schick  have
calculated $^{36)}$ the profiles and from them obtained the saddle-splay modulus
for the internal intefaces. The modulus is positive, thus internal interfaces
in the disordered phase of A,B homopolymers and AB diblock copolymers
are characterized by the negative Gaussian curvature.   

Random copolymers$^{40)}$ are produced by polimerization reaction in the medium
that contains two (or more) distinct monomer species.  The chemical composition
of the copolymers is dictated by a set of bimolecular reaction rates.
The system is characterized by quenched disorder$^{41-43)}$. The  
chain architecture is determined statistically by the conditional
probability that A monomer follows B monomer. Under steady state conditions
of the reaction this probability is independent of the location of the
monomer in the chain. The structures formed in 
the random copolymer system
depend on the structure of the chain and thus on the
polimerization process, thus the phase diagram can 
be a useful guide for the design of new polymeric materials. 

In the multiblock copolymer system$^{44)}$  new ingredients enter into
physics in comparison to diblock copolymers. In the latter the junction 
of the blocks of a given copolymer is constrained to one A/B block
interface; in the former bridging and looping is possible. In the loop
configuration of the AB multi-block copolymer system both junctions can
reside on the same interface (Fig.4). In the bridge configuration the middle
B block spans over one B-rich lamella and joins two A-rich lamellas.
In many material properties the bridge configurations are important$^{45)}$. 
The bridging fraction thus provides important information for the design 
of new materials.         

Ring homopolymers or 
ring diblock copolymers provide yet another example of the influence of chain
architecture on the structure and phase diagram of polymer system.
In the AB ring polymer the A block is connected with the B block
at two junction points, therefore the chain has no free ends. 
The topological constraint introduced by this chain architecture, prevents
the different chains from entangledment or self-knotting (Fig.5). Thus the 
statistics of the rings is not Gaussian and the size of a ring 
is intermediate between
the Gaussian and collapsed size. For the ring homopolymer  in the melt
the scaling arguments$^{46a)}$ 
show that the
radius of gyration scales with the polimerization index, $N$, as $N^{\nu}$,
with $\nu =0.4$. 
The application of RPA to the 
Edwards hamiltonian with the specific constraint of no
knots gives the more accurate value$^{46b)}$ of $\nu=0.445$.
For the ring diblock copolymer the 
computer simulations$^{47)}$ 
give $\nu=0.45$. Please note that at high temperature
the scaling in the melt of ring polymers should be the same as in the
ring diblock copolymer system. The random phase 
approximation, with assumed Gaussian statistics, shows$^{48,49)}$ that
the transition to the ordered lamellar phase occurs at the temperature 
$T_{ring}=1.78 T_{linear}$ (for symmetric rings), where $T_{linear}$
is the temperature for the transition to the lamellar phase in the
linear diblock copolymer system. This result compares very
well with simulations$^{47)}$, despite the fact that the ring statistics  
is certainly not Gaussian.
When we decrease the temperature the rings strongly stretch along the axis 
connecting the centers of mass of two blocks.

So far we have discussed the flexible-flexible diblock copolymer systems,
where the diblock copolymers  are modelled 
as Gaussian chains or, almost equivalently, chains with
freely rotating bonds of fixed length. One can
investigate the changes in the structure and stability of the
disordered (isotropic) phase of the diblock copolymer
system with the stiffness of a copolymer molecule.
Microphase separated block copolymers with blocks of different stiffness
are technologically important as composites. The flexible part of the polymer
provides resistence to fracture while the rigid one resistence to tensile
stresses and thermal stability.      
The simplest system of {\it rigid-flexible} diblock copolymers  
is discussed here$^{50)}$.
In a molecule one part of the copolymer
is a chain with freely rotating bonds 
while the other is the rigid rod. In such a system, the 
density of monomers alone is not enough to specify the structure of the 
rod-rich domains nor the stability limits.
In 
order to have a complete description of the system
we must include a nematic order parameter tensor$^{51)}$.
In the high temperature isotropic phase, the various 
correlation functions between density and  nematic order operators
provide information about the structure. 
 The correlation functions, in the Gaussian approximation, are
found to oscillate with a characteristic period equal to the length of the
rigid part. The lamellar domain boundaries of systems with
longer rods are sharper than those with shorter ones, and the linear
density measured along any direction is anticorrelated with the order
parameter measuring nematic order in that direction. This latter effect is due
to the isotropy of the system, and is relatively independent of the
Maier-Saupe and Flory-Huggins parameters.
This model is a special example of the more general model$^{52)}$ 
shown in Fig.6. Here the A block consists of segments of small size $b_A$
and B block consisting of segments of much larger size $b_B$.     

Liquid crystal polymers and their mixtures are studied both because of their
practical utility and their intrinsic interest$^{53,54)}$. In the case of 
mixtures$^{55,56)}$, one
wants to know how the location of various phases, isotropic and nematic, and
their transitions depend on the properties of the two components, their
rigidities, polymerization indices, interactions etc. An approach to this
problem requires a model for the liquid crystal polymers. Recently, 
Ma\"issa and
Sixou$^{57)}$ addressed these issues on   by modelling the polymers as 
elastic lines with
an energy of bending$^{58)}$. Their  calculation can be applied to
 many different
mixtures, and  they obtained various phase diagrams numerically for several
values of the interaction constants. It is difficult, however, to extract
general relationships between properties of the phase diagrams and the polymer
parameters from their formalism. Such dependences can be made manifest if the
Landau de Gennes expansion of the free energy is derived from the polymer
model because the coefficients in this expansion depend upon the polymer 
characteristics of interest.
Fredrickson and Leibler$^{59)}$, inspired by experiments of 
Moore and Stupp$^{56)}$,  carried out this program in a model
which also treated both polymers as elastic lines, ones characterized by 
different
bending constants. Among other results, they found that the expansion was only
applicable to this model if the two bending rigidities were large and nearly 
equal. On the other extreme there is a model of rigid flexible mixture where
the differences in stiffness are large$^{60)}$.  

There are many phenomena which will not be discussed in this article. 
The mixtures are assumed as incompressible and compressiblity effects are
ignored$^{61)}$. The crossed-linked blends$^{62)}$ and charged polymer 
systems$^{63)}$, although interesting, will not be discussed here. 
We have concentrated in this 
paper on the equilibrium properties of various polymer systems; consequently
the dynamics of blends$^{64-66)}$ and diblock copolymer system$^{67,68)}$
is not included.

The paper is organized as follows: In section II the mesoscopic model of polymer
molecules is presented. Here we discuss the bond configuration for different
systems, the various interaction hamiltonians steming from the Edwards model,
density operators and order parameters and finally the conditional
partition function. In section III we present the detailed study of the 
polymer blend system, including the discussion of the critical temperature
in the Gaussian and one-loop approximation, the single chain statistics and
the problem of the upper wavevector cutoff in polymer blends and in general in
polymer systems. In section IV we discuss the diblock copolymer system.
Here we discuss the behaviour 
of tagged chain in the melt of copolymers, phase diagram etc.
 The system
is studied in the random phase approximation, one-loop approximation and
self consistent field theory. 
In
section V we study the
random copolymer  and multiblock copolymer systems. 
In section VI we present the results for the mixtures of homopolymers
and copolymers.
 The correlations in the rigid-flexible
diblock copolymer system are studied in section VII in the random phase 
approximation. The mixtures of rigid and flexible polymers are
discussed in section VIII. 

\centerline{\bf II. The mesoscopic model of a polymer system}

The mesoscopic model of polymers is specified by the Hamiltonian describing 
the interaction between monomers, and the density distribution of monomers 
within a polymer molecule. Furthermore one has to specify the density operators
which are needed to specify the structure of the system. The average of this
operators is related to the order parameters. Next the conditional
partition function has to be constructed. This is the partition function
for the system subject to the constraint of the the fixed configuration
of the order parameter field. The logarithm of this partition function
expanded in the order parameter gives the Landau-Ginzburg free energy.
The calculations of correlations are also based on this partition function.
Below we specify all the elements of the model. 
\vfill\eject

\centerline{\bf A. Bond configuration}

 The information that N monomers are connected and form a  chain 
 is specified by the 
distribution function $W[{\bf r}]$.
To describe the flexible polymer, we use the 
chain model, in which atoms are described as 
being joined
by freely rotating bonds of fixed length $l$.
The distribution for N atoms in such a chain 
is given by$^{69)}$ 
$$W[{\bf r}]=\displaystyle{\prod_{i=1}^N {{\delta(\vert{\bf u}_i\vert-l)}\over
{4\pi l^2}}}\eqno(2.1)$$
and is normalized as 
$$\int D{\bf r} W[{\bf r}]=1,\eqno(2.2)$$
with
$$D{\bf r}={1\over V}d{\bf r}_0 d{\bf r}_1\cdots d{\bf r}_N.\eqno(2.3)$$
Here ${\bf u}_i={\bf r}_i-{\bf r}_{i-1}$ is a vector 
specifying the orientation of the monomer and $r_i$ is the location of the
$i$ point between the subsequent monomers.     
The same quantitative results at the level of the
Landau-Ginzburg free energy are obtained if instead of the 
previous model we use the Gaussian model in the discrete version
$$W[{\bf r}]\sim\exp\left(-{3\over{2 l^2}}\sum_{i=1}^{N} \left\vert{\bf r}_i-
{\bf r}_{i-1}\right
\vert^2\right)\eqno(2.4)$$
or a continuous version of the model,
$$W[{\bf r}]\sim\exp\left(-{3\over{2 l^2}}\int_{0}^{N} ds 
\left\vert{{d{\bf r}(s)}\over {ds}}\right
\vert^2\right) .\eqno(2.5)$$
In both cases one normalizes the distribution.

The rigid polymers are described as rigid rods, or needles, 
in which all bonds have not only the same length, but also the same
direction. In this case 
$$W[{\bf r}]=\displaystyle{
{{\delta(\vert{\bf u}_1\vert-l)}\over {4\pi l^2}}
\prod_{j=2}^{N}\delta ({\bf u}_{j}-{\bf u}_{j-1})
}.\eqno(2.6)$$

Finally, for a diblock copolymer in which a
flexible chain consisting of N atoms has been joined with a rod
containing M atoms, 
$$W[{\bf r}]=\displaystyle{\prod_{i=1}^N {{\delta(\vert{\bf u}_i\vert-l)}\over
{4\pi l^2}}\prod_{j=N+1}^{N+M}\delta ({\bf u}_{j}-{\bf u}_{j-1})}\eqno(2.7)$$

For semiflexible polymers one can use the model shown on Fig.6 which in the
extreme limit reduces to the one discussed above. A different model of
semiflexible polymers is the wormlike-chain model$^{58,59)}$. In the
discrete version we have:
$$W[{\bf r}]\sim\exp\left({1\over 2}k_1\sum_{i=1}^N {\bf u}_{i+1}{\bf u}_{i}
\right),\eqno(2.8)$$
where $k_1$ is the bending elastic constant, which is the energy penalty for
the change of angle between the nearests monomers.
The continuous version of this model$^{58)}$ is defined as follows:
$$W[{\bf r}]\sim\exp\left(-{1\over 2}k_1\int_0^N ds 
\left\vert{{d{\bf u}(s)}\over {ds}}\right
\vert^2\right),\eqno(2.9)$$ 
subject to the constraint $\vert{\bf u}\vert^2=1$.
The next step in the definition of the model is the specification of the
interactions between the monomers.

\centerline{\bf B. Interactions between monomers}

The typical length scale for a
flexible chain is given  by the radius of gyration
i.e. the size of the region occupied by the chain in the melt.
In the simplest Gaussian approximation one finds that it is proportional to
$\sqrt{N}l$ and is much larger than the monomer size $l$.
All the interesting phenomena take place at the length scale 
proportional to the radius of gyration. On the other hand  the
range of the potential is proportional to $l$, and therefore is
not relevant to the phenomena discussed in the paper. Guided 
by this simple observation the following short range
effective interaction potential has been proposed by Edwards$^{70)}$:
$$v^{\alpha\beta}_{ij}({\bf r}_i,{\bf r}_j)=w_{\alpha\beta}\delta({\bf r}_i-
{\bf r}_j),
\eqno(2.10)$$
where $\delta$ is the Dirac delta function and $w_{\alpha\beta}$    
is the effective interaction parameter for $\alpha$ and $\beta$ type
monomers (e.g. A, B monomers in the binary homopolymer mixture).
The effective interaction parameter is given by the integral$^{70,71)}$ 
of the
direct correlation function$^{72,73)}$.  
In the first approximation for the rigid
molecules, such integral, does not depend on temperature and
 is equal to the excluded volume, the volume 
inaccesible to one molecule when the other is fixed in space. In the case
of two spheres of radius $D$ the excluded volume 
is the larger sphere of radius $2D$. 
Apart from the repulsive forces there is also an attractive potential
e.g. van der Waals potential. The direct correlation function is, in the
first approximation proportional to this potential. Thus $w_{\alpha\beta}$
contains two contribution: one associated with the excluded volume and one
with the attractive potential.
Summarizing: the  interactions 
given by Eq.(2.10) are related to the microscopic interaction potential
via the direct correlation function. The delta function 
signifies the extreme short range character of the potential as 
measured in terms of the length given by the radius of gyration.

The interaction potential for rigid, elongated molecules,
depends not only on the positions but also on the orientations of molecules.
In the case of two rods$^{51)}$ 
with fixed orientations the excluded volume has roughly the
shape of a rectangular box. If we expand the excluded volume in terms 
of the Legendre polynomials of the cosine of the relative
angle of two rods we obtain the constant term, proportionl to
 $w_{\alpha\beta}$,   
plus the second term
related to the second Legendre polynomial, $P_2$. If higher order terms are
neglected the anisotropic
part of the potential is modelled as$^{50,60)}$:
$$v^{\alpha\beta}_{i,j}({\bf r}_i,{\bf u}_i,{\bf r}_j,{\bf u}_j)=
v_{\alpha\beta}
\delta ({\bf r}_i-{\bf r}_j)
P_2\left({{{\bf u}_i
{\bf u}_j}\over {\vert{\bf u}_i
\vert\vert {\bf u}_j\vert}}\right)\eqno(2.11)$$ 
The total potential is given by the sum of Eq(2.10) and (2.11).
Please note that in general the parameters $w_{\alpha\beta}$ and 
$v_{\alpha\beta}$ are not independent, since they follow from the
same direct correlation function.

Finally for the dipolar monomers one can model the potential as:
$$v^{\alpha\beta}_{i,j}({\bf r}_i,{\bf u}_i,{\bf r}_j,{\bf u}_j)=
v_{\alpha\beta}
\delta ({\bf r}_i-{\bf r}_j)
\left({{{\bf u}_i
{\bf u}_j}\over {\vert{\bf u}_i
\vert\vert {\bf u}_j\vert}}\right)\eqno(2.12)$$  

The next step is the specification of the density operators needed for the
definition of the order parameters.  

\centerline{\bf C. Density operators} 

For the sake of clarity we shall discuss this point using 
 the example of the rigid-flexible 
diblock copolymer system$^{50)}$. 
The chain architecture is given by Eq(2.6). The system consists of 
$n$ chains each consisting of $N_A$ A-type monomers joined to
$N_B$, B-type monomers.  The interactions between two monomers 
are described by the sum of 
Eq(2.10) and Eq(2.11). The density operators needed to specify the 
configuration of the system are as follows:
$${\hat \phi}^{(n)}_A({\bf r})={1\over \rho_0}\sum_{\gamma=1}
^{n}\sum_{i=1}^{N_A}\delta({\bf r}-{\bf r}_i^{\gamma});\eqno(2.13)$$ 
$${\hat \phi}^{(n)}_B({\bf r})={1\over \rho_0}\sum_{\gamma=1}
^{n}\sum_{j=N_A+1}^{N_A+N_B}\delta({\bf r}-{\bf r}_j^{\gamma});\eqno(2.14)$$ 
$${\hat Q}^{(A)}_{\alpha\beta}({\bf r})={1\over \rho_0}\sum_{\gamma=1}
^{n}\sum_{i=1}^{N_A}\delta({\bf r}-{\bf r}_i^{\gamma})\left({3\over 2}
{{({\bf u}_i
^\gamma)_\alpha\cdot ({\bf u}_i
^\gamma)_\beta}\over {\vert{\bf u}_i^{\gamma}\vert^2}}-{\delta_{\alpha\beta}
\over 2}\right);\eqno(2.15)$$ 
$${\hat Q}^{(B)}_{\alpha\beta}({\bf r})={1\over \rho_0}\sum_{\gamma=1}
^{n}\sum_{i=N_A+1}^{N_A+N_B}\delta({\bf r}-{\bf r}_i^{\gamma})\left({3\over 2}
{{({\bf u}_i
^\gamma)_\alpha\cdot ({\bf u}_i
^\gamma)_\beta}\over {\vert{\bf u}_i^{\gamma}\vert^2}}-{\delta_{\alpha\beta}
\over 2}\right).\eqno(2.16)$$
The first two operators, $\hat\phi^{(n)}_A({\bf r})$ and 
$\hat\phi^{(n)}_B({\bf r})$ (superscript (n) stands for the number of chains in
the summation), 
represent the microscopic number 
fractions at point ${\bf r}$ 
of A and B monomers respectively;
the next two tensor operators,  $\hat 
Q_{\alpha\beta}^{(A)}({\bf r})$ and 
$\hat Q_{\alpha\beta}^{(B)}({\bf r})$ ($\alpha , \beta =
1,2,3$),
represent the microscopic nematic 
tensor order parameters$^{51)}$ at point ${\bf r}$
for the A and B monomers respectively. These tensors are 
symmetric and 
of zero trace, thus each has five independent components. The total number
of independent operators is, therefore, twelve.  
Here $\rho_0$ is the average density of monomers in the system.

The total interaction hamiltonian, H, 
is obtained by summing all the interactions
between different monomers. Using the  density operators it  
can be rewritten in the following simple form:
 $$\displaystyle{\eqalign{H&=k_BT
\rho_0\int d{\bf r}\Bigr({1\over 2}w_{AA}
(\hat\phi^{(n)}_A)^2({\bf r})+
{1\over 2}w_{BB}(\hat\phi^{(n)}_B)^2({\bf r})+w_{AB}\hat
\phi^{(n)}_A({\bf r})\hat 
\phi^{(n)}_B({\bf r})\cr-{1\over 3} &
v_{AA}\hat Q^{(A)}_{\alpha\beta}({\bf r})\hat Q^{(A)}_{\beta\alpha}({\bf r})-
{1\over 3}v_{BB}\hat Q_{\alpha\beta}^{(B)}({\bf r})
\hat Q_{\beta\alpha}^{(B)}({\bf r})-{2\over 3}v_{AB}
\hat Q_{\alpha\beta}^{(A)}({\bf r})
\hat Q_{\beta\alpha}^{(B)}({\bf r})\Bigr),\cr}}\eqno(2.17)$$
where summation over repeated $\alpha$ and $\beta$ indices is implied.

There is no general recipe for the choice of the density operators. 
The hint as to the right choice is provided 
by the form of the interactions between monomers and the
chain architecture. Here $\hat\phi^{(n)}_\gamma$,
$\hat Q^{(A)}_\alpha\beta$ $\gamma =A,B$
form a minimal set of order parameter operators necessary to describe
the system in the Landau approach.
 
\centerline{\bf D. Conditional partition function}    

The conditional partition function, $Z[P_i]$ 
is the partition function for the system subject to the constraint that
the microscopic operators, $\hat P_i$ be fixed at prescribed
values, $P_i({\bf r})$ i.e.
$$Z[P_i]=\left<\prod_{i}\delta\left(\hat P_{i}-P_{i}
\right)\right>,\eqno(2.18)$$
where the average is calculated as follows:
$$\displaystyle{\Bigr<\cdots\Bigr>={1\over Z_0}
\prod_{\gamma=1}^n\int D{\bf r}^{\gamma}\cdots 
W[{\bf r}^{\gamma}]
\exp(-H/k_BT)}\eqno(2.19)$$
Here $k_BT$ is the Boltzman factor and  
$Z_0$ is the canonical partition function. 
The Landau-Ginzburg (LG) free energy (in the mean-field approximation), 
$\Omega[P_i]$, is given by 
$$\Omega[P_i]=-k_BT\ln{Z[P_i]}.\eqno(2.20)$$
The conditional partition function and the LG free enrgy are functionals
of the order parameters $P_i$. The partition function of the system is
given by the summation of the conditional partional function over all 
possible configurations of the fileds $P_i$:
$$Z=\prod_i\int DP_i Z[P_i].\eqno(2.21)$$    

In the following sections we shall apply this method to polymer systems.

\centerline{\bf III. Binary mixture of flexible homopolymers}

We consider a mixture of n$_A$ A-type polymers with N$_A$ monomers in each 
molecule, and n$_B$ B-type polymers with N$_B$ monomers
in each molecule inside
a volume V. The anisotropic interactions, $v_{\alpha\beta}$ 
are neglected.
For later convenience we introduce two fields,
$$\Psi_A({\bf r})=\phi^{(n_A)}_A({\bf r})-\bar\phi\eqno(3.1)$$ 
and
$$\Psi_B({\bf r})=\phi^{(n_B)}_B({\bf r})-(1-\bar\phi).\eqno(3.2)$$
These fields describe the excess of the fraction
of A or B monomers at the point ${\bf r}$
over their average values in the system,
$\bar\phi=n_AN_A/(n_AN_A+n_BN_B),$ and $1-\bar\phi$;
they are zero in the homogeneous system.
Finally, we assume that the system is incompressible, {\it i.e.}
$${\phi}^{(n_A)}_A({\bf r})+{\phi}^{(n_B)}_B({\bf r})=1,\eqno(3.3)$$
so that 
$$\Psi_A({\bf r})=-\Psi_B({\bf r})=\Psi ({\bf r}).\eqno(3.4)$$
For the Fourier transform of $\Psi$, $\Psi ({\bf q})=\int d{\bf r}
\Psi ({\bf r})e^{i{\bf rq}}$, it implies $\Psi ({\bf q}=0)=0$.  
Since the system is incompressible, only one interaction parameter is
needed, namely the dimensionless Flory-Huggins parameter, $\chi$,
$w_{AA}+w_{BB}-2w_{AB}=-2\chi<0$. 
In general $\chi$ is positive so that it is energetically favorable for 
the
system to separate into A-rich and B-rich phases. 
The first aim 
of the study is the location of the critical point for the mixture,
below which the system separates into A-rich and B-rich phases.

The field $\Psi$ is for the incompressible blend the only order parameter
and consequently both the partition function and the LG free energy are
the functionals of $\Psi$. Please note that $\Psi$ vanishes in the
homogeneous system, but is nonzero at the
coexistance curve. We find, using the results of the cumulant expansion
given in 
Appendix A and B, the following
form of $\Omega[\Psi]$, for $\Psi$ independent on position:
$$\eqalign{
\Omega[\Psi]=&\left({1\over 2}\left({1\over {N_A\bar\phi}}+{1\over {N_B(1-
\bar\phi)}}\right)-{\chi}\right) \Psi^2\cr+&
{1\over 6}\left({1\over {N_B(1-\bar\phi)^2}}-{1\over {N_A\phi^2}}\right)
\Psi^3\cr+&{1\over 12}\left({1\over {N_B(1-\bar\phi)^3}}+{1\over 
{N_A\bar\phi^3}}\right)\Psi^4\cr}.
\eqno(3.5)$$
 Linear terms, in $\Psi$, have been omitted in Eq.(3.5) as they 
simply correspond to a shift, $h_\Psi$, in the exchange chemical potential 
$\mu=(\mu_A/N_A)-(\mu_B/N_B)$, {\it i.e.}
$$h_\Psi=
                  \mu-{\ln\bar\phi\over N_A}+{\ln(1-\bar\phi)\over N_B}-
{1\over N_A}+{1\over N_B}-\chi(1-2\bar\phi).\eqno(3.6)$$
Here $\mu_A, \mu_B$ are the chemical potentials of $A$ and $B$ monomers. 
Please note that Eq(3.5) can be also obtained (as it should) 
directly from the 
Flory-Huggins free energy
by expanding it in $\Psi$. The critical point for the mixture corresponds
to the vanishing of the first and second term in Eq(3.5), or in
other words to the vanishing of the second and third derivative of the
Flory-Huggins free energy with respect to the concentration.  
   The location of the critical point in this mean-field approach (Appendix A
and B) 
for the A,B polymer mixture is given by the well known equations:
$$\displaystyle\chi_c^{mf}={{(\sqrt{N_A}+\sqrt{N_B})^2} 
\over {2N_AN_B}}\eqno(3.7)$$
and
$$\displaystyle\bar\phi_c^{mf}={{\sqrt{N_B}}\over{\sqrt{N_A}+\sqrt{N_B}}},
\eqno(3.8)$$
where 
$\chi_c^{mf}$  is the value of Flory-Huggins parameter at the critical point
and $\bar\phi_c^{mf}$
is the critical concentration of A monomers ($mf$ stands for mean-field). 
In order to include the fluctuations (i.e. dependence of $\Psi$ on
positions) we have to study the correlations functions in the system.
Please note that the $\chi$ parameter is rescaled by $k_BT$.

\centerline{\bf A. Vertex functions and critical point}

First of all, the critical point is marked by a very strong
scattering, i.e. the divergence of the
scattering intensity I(q) at $q\to 0$. This quantity is proportional
to
$$I(q)\sim{({\Gamma_2({\bf q},-{\bf q})})^{-1}}
={1\over V}<\Psi ({\bf q})\Psi (-{\bf q})>.
\eqno(3.9)$$
Here the average is calculated according to the following formula:
$$<\cdots >={1\over Z}\int D\Psi\cdots Z[\Psi ],$$
where $Z$ and $Z(\Psi )$ are defined in section II D.                 

At the critical point, $\Gamma_2$ goes to zero for ${\bf q}\to 0$. 
Additionally,
$$\Gamma_3({\bf q},{\bf k},-{\bf q}-{\bf k})=V^2{{<\Psi({\bf q})\Psi
({\bf k})\Psi (-{\bf q}-{\bf k})>}\over{<\Psi ({\bf q})\Psi(-{\bf q})>
<\Psi ({\bf k})\Psi(-{\bf k})><\Psi ({\bf q}+{\bf k})\Psi(-{\bf q}-{\bf k})>}}
\eqno(3.10)$$
vanishes at the critical point for ${\bf q,k}\to 0$. The two
equations,
$$\Gamma_2(0,0)=0,\eqno(3.11)$$
$$\Gamma_3(0,0,0)=0,\eqno(3.12)$$
locate the critical point in the $(T,\bar\phi )$ plane.

The method dealing with the calculation of
the averages in the form given by Eq.(3.9) 
is described in detail in Ref.[6,7]. 
Here we use an
expansion in the number of loops  (Chapter 6, Ref.[7]), which in the 
case of a polymer mixture has great benefit of being an expansion in 
a small
parameter proportional to $1/\sqrt{N}$. 
In the limit of $N\to\infty$, only
the terms with no loops survive. The zeroth order approximation (no loops)
corresponds to the random phase approximation$^{2)}$ (see Appendix A and B,
Eq.(B.11) and the critical point location is given by Eqs.(3.7-8),
thus it is the same result as the Flory-Huggins result. 

The equations for the vertex functions in the one-loop approximation are
discussed in Appendix C and Fig.7 and Fig.8 show the one loop diagrams which
contribute to the vertex functions.
Now we can proceed with the solution of Eqs.(3.11,12) in the one-loop
approximation (Appendix C). We are interested
only in the first corrections to the mean-field results. Thus, we assume the
solution in the following form:
$$\chi_c=\chi_c^{mf}(1+\alpha_1),\eqno(3.13)$$
and
$$\bar\phi_c=\bar\phi_c^{mf}(1+\alpha_2),\eqno(3.14)$$
where $\alpha_1$ and $\alpha_2$ are small corrections of the order
$1/\sqrt{N}$ and $\chi_c^{mf}$ and $\bar\phi_c^{mf}$ are 
given by Eqs.(3.7-8).In order to simplify the lengthy formulas,
we also assume that the lengths of the monomers are equal,
$l_A=l_B=l$, but, of course, it is 
straighforward to perform the calculations without this assumption.
Combining all these equations we find $\chi_c $ and 
$\bar\phi_c$ in the following form:
$$\eqalign{2\chi_c&={{(\sqrt{N_A}+\sqrt{N_B})^2} 
\over {N_AN_B}}\Biggr(1+\left({{12}\over {\pi\rho_0l^3}}\right)
\left({l\over{\Lambda}}\right)+\left({{4\pi}\over{3\rho_0l^3}}\right)\cr
&\times\left({{N_A^{3/2}+N_B^{3/2}}\over{\sqrt{N_A}+\sqrt{N_B}}}\right)
\left({l\over{\Lambda}}\right)^3-\left({{2\pi}\over{3\rho_0l^3}}\right)
(\sqrt{N_A}-\sqrt{N_B})^2\left({l\over{\Lambda}}\right)^3\Biggr),\cr}
\eqno(3.15)$$
$$\eqalign{\bar\phi_c=&{{\sqrt{N_B}}\over{\sqrt{N_A}+\sqrt{N_B}}}
\Biggr(1+\left({{4\pi}\over{\rho_0l^3}}\right)
{{(\sqrt{N_A}-\sqrt{N_B})^3}\over {\sqrt{N_B}}}
\left({l\over{\Lambda}}\right)^3\cr &+
\left({{36}\over{\pi\rho_0l^3}}\right)
{{(\sqrt{N_A}-\sqrt{N_B})\sqrt{N_A}}\over{(\sqrt{N_A}+\sqrt{N_B})^2}}
\left({l\over{\Lambda}}\right)\cr &+
\left({{4\pi}\over{\rho_0l^3}}\right)
{{(N_A^{3/2}-N_B^{3/2})\sqrt{N_A}}\over
{(\sqrt{N_A}+\sqrt{N_B})^2}}\left({l\over{\Lambda}}\right)^3\Biggr),\cr}
\eqno(3.16)$$
where $\Lambda$ is the cutoff discussed in Appendix D.

For $N_A=N_B=N$, we have $\bar\phi_c=1/2$ and
$$\eqalign{\chi_c&={2\over N}\Biggr(1+\left({{12}\over{\rho_0l^3\pi C_0\sqrt{N}}}
\right)\cr &+\left({{4\pi}\over{3\rho_0l^3C_0^3\sqrt{N}}}\right)\Biggr),\cr}
\eqno(3.17)$$
where $C_0=\sqrt{C_1+C_2}$ is a constant
(see Appendix D). Assuming $C_0=1$ and the 
volume fraction (the ratio of the 
volume occupied by monomers to the total volume)
$\rho_0l^3=1$ we find 
$$\chi_c={2\over N}\left(1+\left({8\over{\sqrt{N}}}\right)\right).\eqno(3.18)$$
We can use this formula to estimate the shift of the critical temperature
induced by fluctuations. Let $T_c^{mf}=400$K for
$N=10^3$. Then we find from Eq.(3.18) that $T_c=319$K. The shift,
$\Delta T=T_c^{mf}-T_c$ is 81 degrees!
Even for $N=10^4$ we have still 8\% correction to $\chi_c/\chi_c^{mf}$.
Of course, the constant $C_0$ should be the fitting parameter in the
analysis of experimental results and its value 
can be larger or smaller than one.
Nonetheless, we think that it should be of the order of 1. 

For $N_A>N_B$, we find that $\bar\phi_c$ is larger than the mean-field
value of the concentration of A-monomers, $\bar\phi_c^{mf}$. For
$N_A\gg N_B$, it follows that
$$\bar\phi_c\approx\sqrt{{{N_B}\over{N_A}}}
\left(1+\left({{4\pi}\over{\rho_0l^3C_1^{3/2}}}\right)\left({1\over{\sqrt{N_B}}}
\right)\right).\eqno(3.19)$$
For $N_A\to\infty$ and $N_B$ finite, the critical concentration
approaches zero, while the critical temperature
approaches $T_c^{mf}$. 

Finally, let us discuss the order of the n-loop diagram in the limit of
$N_A\gg N_B$. In this limit $\bar\phi_c\approx\sqrt{N_B}/\sqrt{N_A}$ 
and is
very small. It enters the n-loop diagram as a multiplicative constant
in the form $1/\bar\phi^{n+1}$. Additionally, we have $1/q^2$ contributions
for $n-1$ loops in the n-loop diagram.
After grouping all these terms and using the formulas for the 
cutoff (Appendix D), we find that any n-loop diagram is proportional to
$1/(\sqrt{N_A}N_B^{(n+1)/2})$, and it follows that the small parameter
in the loop expansion is now $1/\sqrt{N_B}$. Because of the
$1/\sqrt{N_A}$ factor we also find that in the limit of 
$N_A\to\infty$ $T_c\to T_c^{mf}$ even for finite $N_B$. 
This result is in accordance
with de Gennes' predictions$^{1)}$, for the polymer in a solvent.
As we can see, the loop expansion breaks down if either
$N_A$ or $N_B$ is of the order of unity, since then it is no longer
true that the succesive n-loop terms are small in comparison with
the (n-1)-loop terms. This analysis also prove that the loop expansion in polymer 
blends is indeed the expansion in small parameter, providing we chose the
cutoff according to the prescription given in Appendix D.

\centerline{\bf B. The Ginzburg criterion} 
%%%%%%%%%%%%%%%%%%%%%%%%%%%%%%%%%%%%%%%%%%%%%%%%%%%%%%%%%%%%%%%%%%%%%%%%%%%%%%% 
The critical temperature and concentration are the nonuniversal
quantities depending on the details of the system. On the other hand,
the values of the critical exponents are universal and do not depend
on any microscopic details of the system. A set of critical exponents
characterizes the whole class of critical points. For example the
critical exponents for the Ising spin system, low-molecular mass
binary mixtures and polymer binary mixtures are the same. All these
critical points are said to belong to the Ising universality class$^{7)}$.
As was mentioned in the Introduction, the scattering intensity 
diverges on the approach to the critical point, according to the
following formula: $I(q\to 0)\sim (T/T_c-1)^{-\gamma}$. We know that 
for the Ising
universality class $\gamma=1.26$, while in the mean-field 
theory we get $\gamma=1$. 
As we have seen in the previous section the fluctuations shift the
critical temperature, however 
the most important effect of fluctuations
is the change of the value of the critical exponents. In the 
experimental analysis$^{8-13)}$ of $I(q)$, in order to extract
$\gamma$, one plots $\log I(q)$ versus $\log (T/T_c-1)$.
(see Fig.9).
It has been observed that away from the critical point
the plot is a straight line with the slope of 45 degrees, indicating
$\gamma=1$. As one approaches $T_c$ closer the straight line 
begins to curve.
Finally, very close to the critical point it becomes a straight line
again, but with the slope corresponding to $\gamma=1.26$.
The change in the slope
marks the breakdown of the mean-field theory. The criterion that tells us
when this happens is called the Ginzburg criterion. The region around the
critical point where $\gamma\ne 1$ is called the Ginzburg region.
Since $I^{-1}(q)\sim \Gamma_2(q,-q)$, we can base our estimate of the
Ginzburg region on the analysis of the two-body vertex function
given by Eq.(3.9). Our strategy is to write it down as a sum of two terms:
a mean-field one and a correction to it. As we approach the
critical point the corrections grow in comparison to
the mean-field term. Finally they 
are larger than the mean-field term, indicating
the breakdown of the mean-field theory.
This happens inside the Ginzburg region. 

Close to the critical point and in the limit of $q,p\to 0$
we can rewrite Eq.(3.9) and Eq.(3.10) 
in the following form:
$$\Gamma_2(0,0)=A_1(\bar\phi-\bar\phi_c^{mf})^2
+2(\chi_c^{mf}-\chi)+F_1(\bar\phi, \chi),\eqno(3.20)$$
and
$$\Gamma_3(0,0,0)=A_1(\bar\phi-\bar\phi_c^{mf})
+F_2(\bar\phi,\chi),\eqno(3.21)$$
where $A_1=2(1/N_A(\bar\phi_c^{mf})^3+1/N_B(1-\bar\phi_c^{mf})^3)$
and $F_1$ and $F_2$ are the one-loop corrections to
$\Gamma_2(0,0)$ and $\Gamma_3(0,0,0)$, respectively. They have been
discussed in detail in the Appendix C. 
Now we can express $\bar\phi_c^{mf}$ and $\chi_c^{mf}$ in terms of 
the critical concentration $\bar\phi_c$ and the Flory-Huggins parameter
at the critical point $\chi_c$. Now Eq.(3.20) reads as
follows:
$$\Gamma_2(0,0)=A_1(\bar\phi-\bar\phi_c)^2+2(\chi_c-\chi)
+\left(F_1(\bar\phi ,\chi)-F_1(\bar\phi_c,\chi_c)\right)
-2F_2(\bar\phi_c,\chi_c)(\bar\phi-\bar\phi_c).\eqno(3.22)$$
This equation is the basis for the quantitative formulation of the
Ginzburg criterion, namely, the mean-field theory is valid 
providing the first two terms in Eq.(3.22) 
(giving the mean field result for $\gamma$)
are much larger than the last two
terms , which are the corrections. Thus the mean-field theory breaks down
when
$$A_1(\bar\phi-\bar\phi_c)^2+2(\chi_c-\chi)\le
-(F_1(\bar\phi ,\chi)-F_1(\bar\phi_c,\chi_c))+2F_2(\bar\phi_c,\chi_c)
(\bar\phi-\bar\phi_c).\eqno(3.23)$$
Now, we can give the explicit formula for the Ginzburg criterion
in the special case
of $N_A\approx N_B$ and $\bar\phi\approx\bar\phi_c$.
Neglecting the dependence of $\Gamma_4^{(0)}$
on ${\bf k}$, i.e. assuming the approximation
$\Gamma_4^{(0)}({\bf k},-{\bf k},0,0)\approx
\Gamma_4^{(0)}(0,0,0,0)$ we find the 
Ginzburg criterion in following form:
$$2(\chi_c-\chi)+A_1(\bar\phi-\bar\phi_c)^2= 
{{C_g(N_A^{-1}\bar\phi_c^{-3}+N_B^{-1}(1-\bar\phi_c)^{-3})^2}\over {
\rho_0^2(R_A^2N_A\bar\phi_c^{-1}+R_B^2N_B(1-\bar\phi_c)^{-1})^3}}.
\eqno(3.24)$$
Here $C_g$ is a constant, $R_i^2=N_il^2/6$ ($i=A,B$).
For $\bar\phi=\bar\phi_c$, Eq.(3.24) reduces to Eq.(3) of Ref[10] providing
we change $v_m$ to $1/\rho_0$, $\bar\phi_c^{mf}$ to
$\bar\phi_c$, $\chi_c^{mf}$ to $\chi_c$ and $C$ to $C_g/2$.
On Fig.10  we show the plot of the scattering intensity versus the
inverse of the temperature$^{10)}$. Please note that the mean field critical
temperature is not correctly identified on this plot$^{15)}$
(see also$^{74)}$), since we find that
$$I(q\to 0)\sim (\chi_c-\chi)^{-1}\eqno(3.25)$$
where $\chi_c$ {\bf is not the mean-field Flory Huggins }
critical parameter. 
In fact the Ginzburg region is of the order
of $1/N$ whereas the difference between the critical temperature and the
its mean field value goes as $1/\sqrt{N}$. Therefore it is not correct to identify
the mean field critical temperature with the point where the extrapolated
straight line representing $I(q\to 0)$ far from the critical point
crosses the temperature axis.

Finally, we consider the limiting case of $N_A\gg N_B$.
In the first approximation and for
$\bar\phi=\bar\phi_c$, one finds that the mean-field theory breaks
down for 
$${{(\chi_c-\chi )}\over{\chi_c}}\sim{1\over{\sqrt{N_AN_B}}}.\eqno(3.26)$$
Thus, for $N_A\to\infty$, one would expect that the mean-field theory 
is valid in the whole region around the critical point, for any value of
$N_B$ (Note that $1/\sqrt{N_A}$ term here comes from the
$1/\sqrt{N_A}$ term which apppears in the loop expansion).
This result is misleading for the following reasons:
In our analysis we have not specified the temperature below which
we can expect any influence of the critical point
on the behavior of the polymer mixture. Such a temperature is
called the theta temperature and is denoted, $\Theta$. Above this
temperature mixing is strongly favored and the influence of the
critical point on the scattering intensity is negligible. 
Only below $\Theta$ can we expect any sign of critical behavior
(either mean-field or Ising-like). This means that if
the Ginzburg region estimated above is larger than 
the region between $\Theta$ and $T_c$ we will never see the
mean-field behavior, characterized by $\gamma=1$, 
but only the Ising-like behavior, characterized by $\gamma=1.26$.
As shown by Joanny$^{5)}$
$${{T-T_c}\over{\Theta-T_c}}\approx{1\over{\sqrt{N_B}}}\eqno(3.27)$$
marks the breakdown of the mean-field theory in the aforementioned
sense. Finally we see that in the limit of $N_B\to 1$ the mean-field theory
breaks down completely 
even for infinite $N_A$. As mentioned in the previous 
section also the loop expansion breaks down, but 
peculiarly the
critical temperature approaches its mean-field value.        

\centerline{\bf C. Single chain statistics}
 
The single chain statistics is determined in the first place
by the correlation function for the single chain. In the A,B homopolymer
mixture 
we chose one chain (e.g. A) and tag it by coupling it to the fictitious
field $U_A$.  The correlation function is given by $^{17)}$:

$$S_{AA}({\bf q}_1,{\bf q}_2)=
<\hat\phi_A^{(1)}({\bf q}_1)\phi_A^{(1)}({\bf q}_2)>
={{\delta^2 Z[U_A]}\over{Z[U_A]\delta U_A({\bf 
q}_1)\delta U_A({\bf q}_2)}}.\eqno(3.28)$$  
where the partition function (section II) $Z$ 
as a functional of the external field $U_A$ is given by
$$\eqalign{Z[U_A]=&\int D\phi^{(n_A)}_A\int D\phi^{(n_B)}_B 
\exp{(-H_I[\phi^{(n_A)}_A,\phi^{(n_B)}_B])}
\cr &\int DJ_A\int DJ_B\exp\Biggr(i\int {{d{\bf q}}\over{(2\pi)^3}}
\phi^{(n_A)}_AJ_A+i\int {{d{\bf q}}\over{(2\pi)^3}}
\phi^{(n_B)}_BJ_B\cr+&F^{(n_A)}_A[J_A]+F^{(1)}_A[J_A+iU_A]-
F^{(1)}[J_A]+F_B^{(n_B)}[J_B]\Biggr).\cr}\eqno(3.29)$$
Here $F^{(n_\gamma)}_\gamma$ is defined in section II and Appendix A.
Now using this equation at $U_A=0$
we find the following approximate equation for the
single chain structure factor in the incompressible blend:
$$\eqalign{&VS_{AA}({\bf q},-{\bf q})=<\hat\phi^{(1)}_A({\bf q})
\hat\phi_A^{(1)}(-{\bf q})>_0+
{1\over {2\rho_0}}\int{{d{\bf k}}\over{(2\pi)^3}}\cr &
\Biggr(<\hat\phi^{(1)}_A({\bf q})\hat\phi_A^{(1)}(-{\bf q})
\hat\phi^{(1)}_A({\bf k})\hat\phi_A^{(1)}(-{\bf k})>_0\cr &-
<\hat\phi^{(1)}_A({\bf q})\hat\phi_A^{(1)}(-{\bf q})>_0
<\hat\phi^{(1)}_A({\bf k})\hat\phi_A^{(1)}(-{\bf k})>_0\Biggr)
\cr &{\displaystyle{{{\Gamma_2^{(A)}({\bf k})\Gamma_2^{(A)}({\bf k})}
\over
{\Gamma_2({\bf k})}}}}\cr}\eqno(3.30)$$
It includes the ideal term and the first nonvanishing correction to it.
Here $\Gamma_2$ is defined in Appendix C and $\Gamma^{(A)}_2$ in
Appendix A.
As defined in the Appendix A 
the averages with the subscript 0 are taken over the
configurations of
noninteracting system (ideal averages) whereas those without the 
subscript are taken over the configurations of the interacting system.

The equation for the single chain correlation function
in Ref[20] is similar in structure to Eq(3.30), only instead of having
$\Gamma_2^{(A)}({\bf k})\Gamma_2^{(A)}({\bf k})\Gamma_2^{-1}
({\bf k})$ one has there
$<J_A({\bf k})J_A(-{\bf k})>$. The latter is exactly equal to
$\chi(1+2\rho_0\chi \Gamma_2^{-1}({\bf k}))/2$.
Note that if we use the linear approximate relation
between $J_A$ and $\Psi_A$ i.e. $J_A=\Gamma_2^{(A)}\Psi_A$ we would get the 
same result from both approaches. 
This equation for the single chain correlation function should
be supplemented by the self-consistent equations for the
collective structure factor, $\Gamma_2({\bf k})$ (Appendix C).
The equation for the single chain correlation function and the 
collective vertex functions are also coupled via 
the radius of gyration (see Appendix D) given by $^{75)}$:
$$R_A^2={1\over{2N_A^2}}
<\sum_{i=1}^{N_A}\sum_{j=1}^{N_A}({\bf r}_i^{(1)}-{\bf r}_j^{(1)})^2>
\eqno(3.31)$$
It is obtained from Eq(3.30) by differentiating both sides twice
with respect to $q$ and taking the limit of
$q\to 0$. We find 
$$2N_A^2R^2/3=2N_A^2R_0^2/3-{1\over {2\rho_0}}\int{{d{\bf k}}\over{(2\pi)^3}}
f({\bf k})
\Gamma_2^{(A)}({\bf k})\Gamma_2^{(A)}({\bf k})
\Gamma^{-1}_2({\bf k})\eqno(3.32)$$
where $R_0^2=N_Al^2/6$ is the radius of gyration for the ideal chain
(Eq(3.3) with the ideal average) and $f({\bf k})$ is explicitely given in
Appendix E. Since $f({\bf k})$ is always positive we find the first
intuitively obvious result that {\it chains in a blend shrink}
below the Gaussian chain limit.

The role of the cutoff (Appendix D) 
is also clear here. If not for the $\sqrt{N}$ in the
upper wavevector cutoff the correction to the Gaussian
radius of gyration would be nonzero even in
the limit of $N\to\infty$ and thus RPA would not be a 
correct description in this
limit. 

\centerline{\bf IV. Linear flexible diblock copolymer system}

An AB diblock copolymer molecule consists of two blocks (A and B) 
joint together. The blocks are incompatible, but contrary to the case of AB 
homopolymer blend cannot separate 
on the macroscopic scale
due to the chemical bond joining them. Nonetheless as we lower the temperature 
the blocks separate on the scale of the radius of gyration
forming the periodic structures -- the phase transition between the
disordered phase and ordered is called, in the polymer community
jargon, the microphase separation. In the case of the diblock copolymer 
the concentration of A monomers is set by the architecture of the chain,
the average fraction of A monomers is given by 
$$f={{N_A}\over {N}},\eqno(4.1)$$
where $N=N_A+N_B$ is the total number of monomers
in the diblock copolymer molecule. 
The order parameter field,
 $$\Psi({\bf r})=\phi_A^{(n)}({\bf r})-f,\eqno(4.2)$$ 
is defined as
in section II and $n$ is the number of copolymers.
Incompressibility constraint is implied. Contrary to the case of homopolymer 
blends the equilibrium distribution of A monomers at low temperatures,
 given by $\Psi$, is 
not uniform in space.
 
\centerline{\bf  A. Scattering intensity in the random phase approximation}

Leibler$^{22)}$ applied random phase approximation (RPA)
scheme (described 
in Appendix A and B, see also Ref[76,20]) to the system and found the 
following form of the two body correlation function
(Eq(3.9)): 
$$I(q)={{W_1(q)}\over {W_2(q)-2\rho_0\chi W_1(q)}}\eqno(4.3)$$  
where  $W_1(q)=S_{11}S_{22}-S_{12}^2$ and $
W_2(q)=S_{11}+S_{22}+2S_{12}$ depend only on the architecture of the
diblock copolymer (see section II); $W_1$ is the determinat and $W_2$ the 
sum of all the elements of the matrix $S_{\alpha\beta}$, where
$$S_{11}={1\over V}
<\phi^{(n)}_A(q)\phi^{(n)}_A(-q)>_0={N\over \rho_0} g_1(f,x),\eqno(4.4)$$
$$S_{22}={1\over V}<\phi^{(n)}_B(q)\phi^{(n)}_B(-q)>_0={N\over \rho_0}
g_1(1-f,x)\eqno(4.5)$$
and 
$$S_{12}=S_{21}={1\over V}
<\phi^{(n)}_A(q)\phi^{(n)}_B(-q)>_0={N\over {2\rho_0}}
\left(g_1(1,x)-g_1(f,x)-g_1(1-f,x)\right)\eqno(4.6)$$
Here $x=q^2Nl^2/6$, V is the volume and 
$$g_1(f,x)=2(fx+\exp(-fx)-1)/x^2.\eqno(4.7)$$
Please note (compare with 
Appendix A) that since the blocks A and B are joined the
cross corelation function $S_{12}$ is non zero here.
The behavior of $I(q)$ is shown in Fig.11. 
As we increase $\chi$ (lower
the temperature) the peak of $I(q)$ grows and finally diverges
at the well defined (and fixed) $q^*$ vector. Interestingly the 
position of the peak in this approach is fixed independent of
the temperature. For f=1/2  $I(q)$ diverges at $\chi N=10.55$, while for
$f=0.25$ at $\chi N=17.6$. The divergence of the
scattering intensity indicates the onset of ordering at the typical
length scale given by the inverse of the scattering vector $q^*$.  

The order parameter depends on position in the spatially
ordered phase. The free energy in the 
random phase approximation 
has the following form in this case:
$$\Omega[\Psi]=\sum_{n=1}^{\infty}\int {{d{\bf q}_1}\over {(2\pi)^3}}
\cdots {{d{\bf q}_n}\over {(2\pi)^3}}\Gamma^{(0)}_n({\bf q}_1\cdots
{\bf q}_n)\Psi({\bf q}_1)\cdots
\Psi({\bf q}_n),\eqno(4.8)$$
where the vertex functions $\Gamma_n^{(0)}$ are given in Ref[20,76,22]
and the general method of the calculation has been described in Appendix 
A and B.
The equilibrium configuration of the system corresponds to the
minimum of $\Omega$ i.e. to the solution of the functional equation:
$${{\delta\Omega[\Psi]}\over {\delta\Psi}}=0.\eqno(4.9)$$
In practice this complicated equation  is simplified if we assume
a given form (in accordance with the symmetry of the
phase) 
of the spatial dependence of $\Psi({\bf r})$ in the
ordered phase. For example in the lamellar phase:
$$\Psi({\bf r})=A\cos{(q_l z)}.\eqno(4.10)$$    
   Now inserting (4.10) into Eq(4.8), we find the free energy in the
lamellar phase  as a function of two variables, the amplitude, $A$ and the
wavevectors  $q_l$. The period of the lamellar phase is equal to $2\pi/q_l$.
Now the minimum of the free energy in the lamellar phase
 corresponds to the solution of
two algebraic equations:
$${{\partial\Omega(A,q_l)}\over{\partial A}}=0\eqno(4.11)$$
and
$${{\partial\Omega(A,q_l)}\over{\partial q_l}}=0.\eqno(4.12)$$ 
The phase transition between the disordered phase and the lamellar phase 
takes place at the point where both energies are equal.
In the random phase approximation the transition is second order and takes 
place for $q_l=q^*$ and at the point where the scattering intensity
diverges. Apart from the lamellar phase one studies in the same way$^{22)}$,
other phases such as hexagonal phase, bcc phase, obdd phase or others.
For example the simplest approximation of $\Psi$ in the hexagonal phase
has the following form:
$$\Psi({\bf r})=A\sum_{n=1}^3\cos({\bf q}_n {\bf r}_\bot),\eqno(4.13)$$
where $q_1=(1,1/\sqrt{3})q_h$, $q_2=(-1,1/\sqrt{3})q_h$ and $q_3=
(0,2/\sqrt{3})q_h$
are the vectors spanning the first shell (similarly in the lamellar phase 
Eq(4.10)) in the reciprocal space for the
regular hexagonal lattice. 
The form of $\Psi$ given by Eq(4.10) for the lamellar phase and Eq(4.13) for the
hexagonal phase is a good approximation close to the weakly first order
or second order phase transitions, it breaks down at low temperatures,
when the ordered domains are sharp$^{76)}$.   

 In the random phase approximation the hexagonal phase and bcc phases are
also stable in some range of parameters. The transitions between these
phases and disordered phase are first order. The phase diagram of the diblock
copolymer system is shown in Fig.12.
In the weak segregation regime
the random phase approximation is exact in the limit of $N\to\infty$.  

\centerline{\bf B. Fluctuations induced first order phase transition}

The transition to the lamellar phase from the disordered phase is second
order in the random phase approximation. 
It can be easily seen if we use Eq(4.10), apply it to Eq(4.8)
and expand the free energy in the amplitude $A$. We find the following
form of $\Omega$ (keeping the first three terms only):
$$\Omega = aA^2+bA^4+cA^6\eqno(4.14)$$
Here $b$ and $c$ are two positive constants independent of the temperature 
and $a$ changes sign at the point when
the scattering intensity $I(q)$ (Eq(4.3)) diverges. At this point we have
the second order phase transition. Now if we go beyond
the random phase approximation and include the one-loop corrections the
second term in Eq(4.14) becomes negative before $a$ does, but $c$ remains 
positive 
and consequently the transition becomes first order$^{20,23,77)}$. The
free energy in the one loop approximation has the same general form
given by Eq(4.8), but the coefficients in the expansion are given by the
vertex functions $\Gamma_n$ calculated in the one loop approximation.
The equation for the two and three body vertex function in the one-loop
approximation is given in Appendix D. These equations are general and
valid for polymer blend system as well as for the
disordered phase of diblock copolymers. 
In the ordered lamellar phase the equation for
vertex function are slightly different, since the field $\Psi$ given by
Eq(4.10) is
non-uniform.   
In the case of $f=1/2$ $\Gamma_3=0$ due to the symmetry$^{22)}$ and
the one-loop equation for the  lamellar phase has the form
given in Appendix F$^{20)}$.
The phase diagram in the one-loop aproximation is shown in Fig.13.  
As we can see the transition to the lamellar phase occurs not only
for the symetric case $f=1/2$ (as we have found in the random
phase approximation), but also in the neighbourhood of this
point. The transition to the lamellar phase is first order
as mentioned previously. In the limit of $N\to\infty$ the 
one-loop corrections vanish and the results reduce to the 
random phase approximation.

In the random phase approximation the peak in the scattering intensity
is fixed and scales with the polimerization index as 
$\sqrt{N}$.
However from the experiments$^{78)}$ it follows that the peak in
the structure factor exibits a much stronger
dependence on $N$ than predicted by RPA. In Fig.14 the 
comparison of the results of the one-loop
approximation$^{20)}$ and experiment are shown. The
logarithm of the peak position is plotted against the 
logarithm of N. 
The agreement between the theory and experiment is remarkable, especially
that no adjustable parameters has been used. The vertical straight line
is the predicted N for which the microphase separation occurs (the
temperature is fixed here at T=296 K). Close to this transition
the peak position $q^*$ is approximately represented by
$N^{-0.8}$.
 
Since the peak position deviates strongly from the $\sqrt{N}$ behavior,
it indicates that the conformation of a single polymer molecule 
should also differ from the Gaussian behavior. Indeed, both in the
theory$^{20)}$ and computer simulations$^{21)}$, strong deviation from the
Gaussian behavior has been found. The equations for the radius of gyration
for the polymer blend are discussed in section III, the very similar
equations are found for the diblocks$^{20)}$. The solution of these
equations indicate that the chains strongly stretch as the temperature is
lowered towards the microphase separation transition temperature.
The configuration of the chains indicates that stretching occurs 
at the point where the two blocks are joined. However the blocks 
themselves shrink below the gaussian limit. Thus the configuration of the 
chain resembles the dumbell,  consisting of two coils
more densely packed than the Gaussian coils, and stretched above the
Gaussian limit. The fact that the coils shrink below the Gaussian limit
has the same origin as the shrinking of the polymer chains in the blends
close to the critical point (section III C). Interestingly the deviations
from the Gaussian behavior occur far from the microphase separation
temperature, deep in the disordered phase.

\centerline{\bf C. Gyroid phase and self consistent field theory}

The diblock copolymer system forms a variety of phases. So far we have
only mentioned the simplest structures studied in the theory: the
lamellar, hexagonal and body center cubic structures. Apart from those
simple morphologies the diblock copolymer system exhibit new type of
structures: the bicontinous phases where the junction between the
blocks are located at the periodic continuous internal interfaces.
The double diamond  and gyroid structures are shown in Fig.1 and 2.
Both were observed in the polystyrene-polisoprene system. The first
one in the strong segregation regime$^{30-32)}$ (which is not discussed in this
paper),while the second one in the weak segregation regime$^{79)}$, where
the Landau theory should be valid. 
This phase has not been studied in the random phase approximation, but
in the mean-field theory called the self consistent field
theory$^{24)}$. The latter theory applied to the system of diblock
copolymers predicts the stable gyroid phase between the lamellar and
hexagonal phase as shown on Fig.15. In this theory the problem
of computing the partition function for n chains is reduced to
the single chain in the external field. The difference between 
this approach and the
random phase approximation starts at the level of the partition function
for the system of noninteracting chains in the external fields
$J_A$, $J_B$(Appendix A Eqs(A.7-8)). 
In the calculations this function is computed exactly in the self
consistent field theory, 
while in the random phase approximation it is
expanded in $\Psi$. Both theories neglect fluctuations. The 
gyroid phase ha not been studied  
in the random phase approximation. 

The gyroid structure
is stable for $\chi N$ between 14 and 20, which is in agreement with
experiment. Please note that for $\chi N$ larger than 20 the double
diamond structure is stable (see Fig.3). There is no direct transition
between the disordered and gyroid structures. Instead at $f=0.452$ and
$\chi N=11.14$ there is a triple point, where the lamellar, hexagonal
and gyroid phase coexist. Below $\chi N=11.4$ the hexagonal phase transforms
directly into the lamellar phase. In this approach the double diamond (obdd)
structure is only metastable, but as we mentioned previously it
is understable since the theory is applied to the weak segregation
regime, while the obdd phase is expected at large $\chi N$ in the
strong segregation regime.

The same method has been applied to the diblock copolymer system
with conformational asymmetry, i.e. to the system where the
length of the monomers $l$ is different for different types of
monomers. The phase diagram in the case of $l_A/l_B=\sqrt{10}$ is 
show in Fig.16. By comparing the diagram in Fig.15 ($l_A/l_B=1$)
and Fig.16
we note that conformational asymmetry does not change the 
topology of the phase diagram$^{25)}$.

\centerline{\bf V. Other flexible block copolymer systems}

The variety of different block copolymer architectures is infinite.
Therefore, here we shall chose only a few, to prove that the
methods discussed so far can also be applied to block copolymers
irrespective of the specific architecture. Here we discuss,
in the random phase approximation, 
the triblock copolymers, diblock ring copolymers and random copolymers.

\centerline {\bf A. Ring copolymers}

The AB diblock ring copolymer consists of two blocks joined in two
points. The ring architeture can be easily included in the general
scheme presented in section II. The same method as applied to
polymer blends or linear diblock copolymers and described in Appendix A, B
and C
can be  also applied
to diblock ring copolymers. Here we consider the scattering intensity and
the instability (same as in section IV A) in the system.

The scattering intensity $I(q)$ for ring copolymers, calculated
in RPA, has exactly the same
formal structure as given for linear diblock copolymers in section IV A
(Eq.(4.3)).
Only the matrix $S_{\alpha\beta}$ is different$^{48)}$, i.e.
$$S_{11}={N\over\rho_0}\int_0^f ds\int_0^f ds'{\rm e}^{(-x\vert
s-s'\vert)(1-\vert s-s'\vert)},\eqno(5.1)$$ 
$$S_{22}={N\over\rho_0}\int_f^1 ds\int_f^1 ds'{\rm e}^{(-x\vert
s-s'\vert)(1-\vert s-s'\vert)},\eqno(5.2)$$ 
$$S_{12}={N\over\rho_0}\int_0^f ds\int_f^1 ds'{\rm e}^{(-x\vert
s-s'\vert)(1-\vert s-s'\vert)},\eqno(5.3)$$
$x$ is defined after Eq(4.7).
The same formulas are obtained for the linear diblock copolymers 
but without the
term $1-\vert s-s'\vert$ in the integrand.
 Due to this term
the integrals cannot be calculated analytically for ring copolymers.
 
Let us first compare the scattering intensity of the linear
and ring diblock copolymers in the case of $\chi=0$ 
(high temperature) and $f=1/2$ (symmetric case).
In Fig.17 the scattering intensities are shown. The
peak in the ring copolymer is at larger q vector (which could be expected)
and is less intense than the peak for the liner diblocks.
In Fig.18 the instability line is shown (at which the scattering
intensity diverges). As we see the rings require a larger amount
of repulsion between A and B monomers to segregate. The latter result is
also understandable since the block in the ring are more constrained
than in the diblock. To see how severe are these constraints it is
instructive to compare the instability in the ring copolymer to that
in the $(AB)_n$ copolymer$^{80)}$ 
where there are n AB diblocks joined together
in the linear chain. The size of the AB diblock is N and each
blocks has the size N/2 (symmetric case). It appears that in the
limit of $n\to\infty$ the critical Flory Huggins parameter 
approaches a limiting value $\chi N\to 15$, whereas for the
symmetric diblock rings (see Fig.18) of size N the instability occurs
at $\chi N=17.8$, thus the transition occurs at even
lower temperature. 
In the diblocks the same instability occurs at
$\chi N=10.5$, thus the transition temperature to the lamellar
phase is approximately 40 \% lower in the ring system than in the
linear diblock copolymer system. 

The result for the microphase separation obtained in the random phase
approximation has been confirmed in the computer simulations$^{47)}$.
This result is in fact surprising, since in the random phase approximation
it is assumed that chains adopt the Gaussian configurations, whereas
in the ring system the constraints are such that the statistics is
not Gaussian even in the limit of $N\to\infty$. The scaling exponent for
the radius of gyration is $\nu =0.45$ ($R_g\sim N^{\nu}$) as has been
found in computer simulations$^{47,81,82}$ and theory$^{46b)}$.
The comparison of the radius of gyration for the
linear diblock chain  and ring diblock chain is shown in Fig.19.
Here $T_c$ is the transiton temperature to the lamellar phase 
and roughly concides with the instability temperature shown in Fig.18.

We have learned that the linear diblock copolymer system stretch as the
temperature is lowered$^{20,21)}$. The same applies to ring diblock
copolymers$^{47)}$. As the temperature is lowered the strong increase
(20-25\%) 
in the distance between the center of masses of two blocks is observed
(Fig.20). At the same time the blocks themselves shrink (Fig.21) similarly
as in the case of linear diblock copolymer system discussed in section IV.
The effect of the stretching is more pronounced in the rings
(20-25\% above the radius of gyration at the infinite
temperature) than in the linear diblocks (15\% above the high temperature
radius of gyration). 

\centerline{\bf B. Triblock copolymers}

The ABA triblock copolymer system has been studied  by several
authors$^{83-86)}$ 
The phase diagram for this system is 
similar to the one observed in diblock co\-po\-ly\-mers$^{84,85)}$ 
(Fig.22).
This observation has been supported by experiments$^{83,87-89)}$.
This is not surprising since cutting the triblock in the middle
produces two diblock coplymers and the entropy gained from doing this
should not be very large, so the free energy of triblocks 
in a particular phase should be roughy the same as that of twice as many
diblocks of half the length in their analogous phase.

The scattering intensity for the triblock copolymer system has the 
same general form as given by Eq(4.3), only the matrix $S_{\alpha\beta}$
is different$^{83)}$ ($g_1$ is defined in Eq(4.7))i.e.
$$\eqalign{S_{11}=&{N\over{\rho_0}}
\Bigr(g_1(f_1,N)+g_1(f_2,N)+g_1(f_3,N)\cr+&g_1(1,N)-
g_1(1-f_3,N)-g_1(1-f_1,N)\Bigr)\cr},\eqno(5.4)$$
$$S_{22}={N\over{\rho_0}}g_1(f_2,N),\eqno(5.5)$$
$$\eqalign{S_{12}=&{N\over{2\rho_0}}
\Bigr(g_1(1-f_1,N)+g_1(1-f_3,N)+g_1(1,N)\cr-&
g_1(f_3,N)-g_1(f_1,N)-2g_1(f_2,N)\Bigr)\cr},\eqno(5.6)$$
Here $f_1,f_2,f_3$ are the monomer fractions in each block, $1,3$ are
A blocks.In the case of $f_3=0$ (or $f_1=0$) the 
formula reduces to the diblock copolymer formula given in Eqs(4.4-4.6).

We can also expect that as we lower the temperature the 
chains stretch, but the individual blocks shrink.

There is one notable difference between the system of ABA triblocks
and AB diblocks. In the former case there are bridge 
and loop configurations
(Fig.4), depending on the location, in lamellas, 
of the different A blocks belonging to 
the same triblock. The presences of bridges should affect the mechanical 
properties of the triblock system$^{86)}$ or in general the 
multi-block copolymer system$^{44)}$. It has been found that
about 40-50 \% of copolymers form bridges in the triblock copolymer
lamellar phase.

\centerline{\bf C. Random copolymers}

In the AB random copolymer system$^{40)}$ the architecture 
of the chain is specified not only by the fraction of
A monomers in the chain, $f$, but also by the 
conditional probabilities that the A monomer in the 
chain is followed by the A, $p_{AA}$, or B monomer, $p_{AB}$ and similarly
for B monomers. One naturaly has $p_{AA}+p_{AB}=1$ and same for B.
The function $f$ and
$\lambda=p_{AA}+p_{BB}-1$ are sufficient to describe the disorder quenched
into an AB random copolymer.
Now apart from the densities already introduced in section II we have to
include the occupation number $\theta_i$ at the i-th point 
along the chains. As in all the systems with quenched disorder  one has to
calculate all the relevant thermodynamic quantities for fixed distribution
of $\theta$ and then average them with respect to $\theta$, which
describes the disorder. The practical way to do it is by the replica
method$^{40-43,90)}$. 

Let us study the macrophase separation in the melt, which is
similar to the separation in A, B homopolymer blend.
The condition for the instability in the melt
in the random phase approximation is:
$$4f(1-f)\chi_c^{mf}={{2N(1-\lambda)^2}\over{N(-\lambda)^2+2\lambda\left(
N(1-\lambda)+\lambda^N-1\right)}}.\eqno(5.7)$$
For $\lambda\to 1$ this condition reduces to the same equation as
we would get for the mixture of A homopolymers and B homopolymer
each of length $Nl$ i.e.
$$4f(1-f)\chi_c^{mf}={2\over N}.\eqno(5.8)$$
In the limit of $\lambda\to 0$ we would have a
system where the copolymers have the very short sequences of
the A and B monomers. Thus the condition (5.7) reduces to
$$4f(1-f)\chi_c^{mf}=2\eqno(5.9)$$
as for the system of disconnected monomers.
As we see in this limit the interesting scale of the phenomena 
is related to the short sequence of the A or B monomers and 
not to the size of the polymer molecule.
Consequently our description is no longer valid; in particular 
the structure and phase transitions in the system depend
on the details of the intermolecular potential, which in the present
treatment have been neglected.

\centerline{\bf VI. Mixtures of homopolymers and diblock copolymers}

So far we have considered the polymer melts or blends 
which were described
by a single order parameter. 
In order to illustrate the application of the method presented here
to more complicated
polymer blends 
we consider a ternary mixture of A, B homopolymers and AB diblock
copolymers$^{34-37)}$. 
This system is in some respects similar to the ternary mixture of 
oil water and amphiphiles$^{38,91)}$. 
The copolymer simply acts as 
a surfactant accumulating at the interface of A-homopolymer rich phase and
B-homopolymer rich phase, reducing in this way the number of unfavorable
contacts between A and B monomers$^{36,92)}$. 
This analogy will be exploited here.

\centerline{\bf A. Structure factor matrix in the RPA}
  
To simplify the calculations we assume that the length of all
the polymers is N and the diblocks are symmetric.
It is convenient to introduce a set of three order parameters.
The first order parameter, $\eta$, measures the deviation from its average 
value of the 
concentration difference of A and B homopolymer.
The second $\varepsilon$, measures the local deviation from its average 
value of 
the copolymer concentration. The third, $\Psi$, measures the local
deviation from its average value of the total fraction of A-monomers.
All order parameters vanish in a spatially uniform,
one phase region.

Now the scattering intensity 
$$I(q)=\left(\Gamma_2^{(0)}(q)\right)^{-1}\eqno(6.1)$$ 
becomes a matrix
and so does the vertex function $\Gamma_2^{(0)}$. 
For the matrix
$$\Gamma _2^{(0)}=
\left\vert\matrix{\Gamma _{\eta\eta}&\Gamma _{\eta\varepsilon}&\Gamma _
{\eta\Psi}\cr
\Gamma _{\eta\varepsilon}&\Gamma_{\varepsilon\varepsilon}&0\cr
\Gamma _{\eta\Psi}&0&\Gamma _{\Psi\Psi}\cr}\right\vert,
\eqno(6.2)$$
calculated in the random phase approximation (Appendix A,B)
we obtain
$$\Gamma _{\eta\eta}={1\over 4}\Biggr ({1\over {S_A}}+{1\over {S_B}}+
{2\over {S_{AA}-S_{AB}}}\Biggr ),\eqno(6.3)$$
$$\Gamma _{\eta\varepsilon}={1\over 4}\Biggr ({1\over {S_A}}-{1\over {S_B}}
\Biggr ),\eqno(6.4)$$
$$\Gamma _{\eta\Psi}=-\Biggr({1\over {S_{AA}-S_{AB}}}\Biggr),\eqno(6.5)$$
$$\Gamma _{\varepsilon\varepsilon}= 
{1\over 4}\Biggr ({1\over {S_A}}+{1\over {S_B}}+
{2\over {S_{AA}+S_{AB}}}\Biggr ),\eqno(6.6)$$
$$\Gamma _{\Psi\Psi}= {2\over {S_{AA}-S_{AB}}}-2\chi,\eqno(6.7)$$
where
$$S_A={{(1-\phi)}\over {2\rho_0}} Ng_1(1,x),\eqno(6.8)$$
$$S_B={{(1-\phi)}\over {2\rho_0}} Ng_1(1,x),\eqno(6.9)$$
$$S_{AA}={{\phi}\over{\rho_0}} Ng_1(1/2,x),\eqno(6.10)$$
$$S_{AB}={\phi\over {2\rho_0}}N(g_1(1,x)-2g_1(1/2,x)),\eqno(6.11)$$
and $\phi$ is the average concentration of copolymer.
For simplicity we have assumed that
the concentrations of homopolymers A and 
homopolymers B are equal.
Function $g_1$ is given by Eq(4.7).

The two most useful scattering intensities,
are obtained by taking the inverse of the vertex function matrix Eq(6.1),
and are given below:
$$I_{\Psi\Psi}={(S_A+S_{AA}-S_{AB})\over 
2[1-\rho_0\chi(S_A+S_{AA}-S_{AB})]},\eqno(6.12)$$
in agreement with Broseta and Fredrickson$^{34)}$
and
$$I_{\eta\eta}={2S_A[1-\chi(S_{AA}-S_{AB})]\over[
1-\rho_0\chi(S_A+S_{AA}-S_{AB})]},\eqno(6.13)$$
in agreement with Leibler$^{92)}$.

For completeness, we compute also the copolymer copolymer structure 
function at equal homopolymer concentrations,
$$\eqalignno{I_{\varepsilon\varepsilon }&=2\bigl({1\over
S_A}+{1\over
S_{AA}+S_{AB}}\bigr)^{-1}\cr
&=Ng_1(1,x)\phi(1-\phi)/\rho_0.&(6.14)\cr}$$ 
This result, in agreement with Leibler$^{92)}$, 
is independent of temperature, and 
always has a peak only at zero wavevector only. In 
this respect it is similar to calculated amphiphile-amphiphile 
structure functions in systems of oil, water and surfactant$^{38,94)}$.

\centerline{\bf B. Stability limits, disorder, Lifshitz and equimaxima lines}

Much of the phase diagram can be obtained from either of the above structure 
functions. The limit of stability of the disordered phase occurs at the 
highest temperature at which a 
divergence in the above structure functions appears. 
From Eqs. (6.12) or (6.13), we see that this happens at the smallest 
value of $\chi$. Here we treat $\chi$ as a function of $x$ so that
we can minimize it with respect to $x$. From the denominator of
the structure functions we find
$${1\over \rho_0\chi(x) N}={1\over N}(S_A(x)+S_{AA}(x)-S_{AB}(x)),\eqno(6.15)$$
or, using the explicit formulas,
$${1\over\chi(x) 
N}={e^{-x}(1-2\phi)+4\phi e^{-x/2}+x-1-2\phi\over x^2},\eqno(6.16)$$
where again, $x=q^2R^2\equiv q^2N^2l^2/6$. Minimizing $\chi(x)$, 
we find that for 
$\phi\le 2/3$, the minimum occurs at $x=0$, and that its value is
$$\chi N={2\over 1-\phi},\qquad \phi\le 2/3.\eqno(6.17)$$ 
Along this line, the disordered phase makes a 
continuous transition to uniform ($q=0$) phases. This is just the macroscopic 
phase separation into A-rich and B-rich phases. For a copolymer density 
between $2/3$ and 1, the instability occurs at a non-zero value of $x$
and the continuous transition is to a phase 
characterized by the corresponding non-zero wavevector
$(x)^{1/2}/R$; this is the microphase separation to a lamellar phase.
The density of copolymer at which a transition occurs
 to a phase characterized by $x\neq 0$ is
$$\phi(x)={-2+x+(x+2)e^{-x}\over 
2[2+(x+2)e^{-x}-(x+4)e^{-x/2}]},\qquad 0<x\leq 3.785\eqno(6.18)$$
and the value of $\chi N$ at which it occurs is given by Eqs. (6.16) 
with $x$ in the range indicated in Eq. (6.18).
The transition lines are shown in Fig. 23. There 
is a region of three phase coexistence between the A-rich, B-rich and lamellar 
phases which is shown schematically in the figure. 

Much more information about the disordered phase can be extracted from the 
structure functions, of course. To obtain some information on how the 
copolymer tends to order the homopolymer and also to order itself, it is 
useful to introduce the disorder line and the Lifshitz
lines$^{35,38,39)}$. The disorder 
line is the locus of points in the phase diagram at which a general 
correlation function, $G({\bf r})$, no longer decays monotonically with 
distance at large $r$, but acquires a component which, at large distances, 
behaves like an exponentially decaying {\it oscillatory}
function;
$$G({\bf r})\rightarrow{\exp (-r/\xi)\over r}
\sin(2\pi r/\lambda+\delta).\eqno(6.19)$$
This oscillatory behavior reflects the tendency of the copolymer to order the 
A and B monomers in space. Thus, on the copolymer-rich side of the 
disorder line, 
two lengths are needed to describe the disordered phase; the usual correlation 
length $\xi$, and a wavelength $\lambda$. As the amount of copolymer in the system is 
increased, one expects the wavelength to decrease as order can be 
enforced over shorter distances; conversely, as the copolymer density 
decreases, $\lambda$ increases and diverges at the disorder line.
We locate the disorder line as follows. The Fourier transform of a
general correlation function can be written as a linear combination of the 
$I_{ij}(q)$, and the poles of these functions determine the behavior of the 
correlation function. If these poles are located at a value of $q$ which is 
solely imaginary, the correlation 
function decays monotonically with distance. Thus the disorder line is the locus at which 
these poles first acquire a non-zero real part. As $I_{ij}=(\Gamma_2^{(0)}
)^{-1}_{ij}$, the location 
of these poles is also the locus 
at which the determinant $\vert\Gamma_2^{(0)}(q)\vert$ vanishes
for imaginary $q$. Because this determinant is common to all $I _{ij}$, 
all correlation functions have the same disorder 
line in general. From the structure functions $I_{\Psi\Psi}$ or 
$I_{\eta\eta}$ 
we see that
the disorder line is found by locating the poles of either correlation 
function, just as the phase boundary was located. However 
to locate the phase boundary,
we sought 
the poles occurring for {\it real} $q$ (corresponding to $x\geq0$), whereas to 
locate the disorder line
we seek the poles occurring for $q$ just as it becomes 
purely {\it imaginary} (corresponding to $x\leq 0$). We do this by setting
$x=-y+i\delta$, $y\geq 0$ in Eq. (6.16), and equating real and imaginary parts. 
Then taking the limit of $\delta$ to zero, we obtain two equations which can be 
solved for $\phi(y)$ and $\chi(y) N$ on the disorder line. They are
$$\phi(y)={(e^y(2-y)-y-2)\over2[e^y(2-y)-e^{y/2}(4-y)+2]},\eqno(6.20)$$
and
$${1\over \chi(y) N}={e^{3y/2}-e^y(y+1)+e^{y/2}(y-1)+1\over 
y[e^{y/2}(4-y)-2-e^y(2-y)]},\eqno(6.21)$$
where $y$ ranges over all positive real numbers. Two limits readily emerge 
from these equations. They are that $\phi$ approaches $2/3$ as $\chi N$ 
approaches 6, and that $\phi$ approaches $1/2$ as $\chi N$ vanishes.
The disorder line is shown 
in the phase diagram of Fig.24 as the line LD.

 The disorder line shows where oscillatory behavior first appears in the 
correlation functions of the disordered system. The locus at which this 
behavior first {\it dominates} a particular correlation function is given 
by one of two kinds of line. The first is the
Lifshitz line; at this line, the peak in the associated structure function 
moves continuously from zero to a non-zero wavevector. Its location is 
determined by
$${\partial^2I_{ij}({\bf q})\over \partial q^2}\bigr|_{q=0}=0.\eqno(6.22)$$
In general, each structure function has its own, distinct, Lifshitz line, and 
the difference in these lines provides information on the different effects 
the copolymer has on the various components of the system. The Lifshitz line 
associated with the structure function of all A monomers,  
is readily found to be given by $\phi=2/3$, independent of 
temperature. This line is shown as LF in Fig. 24. 
The fact that the disorder 
line and the Lifshitz line of $I_{\Psi\Psi}$ are close indicates that once 
the tendency of the copolymers to order the A monomers appears in the 
correlation functions, relatively few additional copolymers are needed to make 
this tendency dominant in the total A, total A correlation function. It is not 
clear, however, whether the copolymer affects the A monomers in the copolymers 
or in the homopolymers more strongly, or whether it affects them equally. This 
can be determined by examining $I_{\eta\eta}$, which depends only on the 
homopolymer concentration, and finding the locus at which this function is 
dominated by components of non-zero wavenumber. In the vicinity of the point 
$\chi N=6$, $\phi=2/3$, this dominance is seen as the peak at zero wavevector 
moves to nonzero q, a movement which occurs at the Lifshitz line of this 
structure function;
$${1\over \chi N}=\bigl[{(1-\phi)\phi\over 8}\bigr]^{1/2},\eqno(6.23)$$
However, another mechanism takes over beyond $\phi\approx 0.79$.
One finds 
instead that a second peak at non-zero q develops in $I_{\eta\eta}$, and 
becomes larger than the peak at $q=0$. This occurs before the Lifshitz line 
condition, Eq. (6.22), is satisfied. This new line, which we refer to as an 
equimaxima line, is defined as the locus of points at which the two peaks,
one at zero, the other at non-zero wavenumber, are of equal height. At this 
line, the wavevector characterizing which is the largest peak in the structure 
function moves discontinuously from zero.
The line is determined by the simultaneous solution of
$${\partial I_{\eta\eta}(x,\phi,\chi)\over \partial x}\big|_{x^*}=0,
\eqno(6.24),$$
and
$$I_{\eta\eta}(x^*,\phi,\chi)=I_{\eta\eta}(0,\phi,\chi).\eqno(6.25)$$
The equimaxima line (EG on Fig. 24) and Lifshitz line (LE) join smoothly 
as $x^*\rightarrow 0$ which occurs at $\phi=121/153\approx 0.79$ (point E).
We clearly see from Fig.24 that even though oscillatory components enter all
correlation functions at the disorder line and dominate the correlation 
function of {\it all} A 
monomers with only a rather small additional increase of 
copolymer concentration, they do not dominate the correlation 
function of those monomers {\it only} in the homopolymers 
until a large increase of copolymer concentration,
so large that the system is about to become unstable to the lamellar phase.
We conclude, therefore, that the copolymer is very efficient in self 
organization, but very inefficient in the organization of homopolymers.

\centerline{\bf C. The Landau-Ginzburg free energy and the Lifshitz point}

To simplify the calculations we consider the case of equal 
homopolymer concentrations and integarte out the
concentration of copolymers $\varepsilon$. Therefore 
we are left with one single order parameter $\eta$.
We 
find a Landau free energy in the following form:
$$\Omega[\eta ]={\displaystyle{{1\over 2}}}
k_BT\rho_0\int d{\bf r}\Biggr (h_0\vert \nabla^2\eta 
({\bf r})\vert^2+g_0\vert\nabla\eta ({\bf r})\vert^2+
f(\eta)\Biggr)\eqno(6.26)$$
where the bulk free energy is:
$$f(\eta)=a\eta ^2+
b\eta ^4+c\eta ^6.\eqno(6.27)$$
Here $g_0$ and $h_0$ are constants indepent on
$\eta$, but in general they do depend$^{36)}$ on $\eta$ ($h(\eta=0)=h_0$
and $g(\eta=0)=g_0$).
The parameters $a,b,c$ and $h_0,g_0$ are given in the Appendix G.
An ordinary critical point occurs when $a=0$, while the other parameters are
positive, and an ordinary
tricritical point occurs when $a=0$ and $b=0$, with the other parameters 
positive. Because $g_0$ is positive, the ordered states which occur just below 
these transitions are spatially uniform, characterized by a vanishing 
wavevector $q$. At a Lifshitz point, the stability of these uniform states is 
lost because $g_0=0$ there, so that these points mark the last appearance of 
uniform ordered states and the first appearance of non-uniform ordered states
 (e.g. a lamellar 
one) characterized by a non-zero wavevector. At a Lifshitz critical point, 
$a=g_0=0$; similarly at a Lifshitz tricritical point, $a=b=g_0=0$, with all 
other parameters positive. One sees immediately how unusual this point is 
expected to be as three coefficients must manage to vanish simultaneously.
As you can see from the results given in Appendix G the point L
on Fig.24 is the Lifshitz tricritical point.

 We first determine the behavior of the uniform solution, $\eta_0$, as the 
Lifshitz tricritical point located at $T_c$ is approached ($g_0=b=0$, $a\sim 
t\equiv (T-T_c)/T_c)$ by variation of the free energy of Eq(1.1) and find 
$\eta_0\sim t^\beta$ with $\beta=1/4$. Inserting this into Eq. (6.26), we find
$F\sim t^{2-\alpha}$ with $\alpha=1/2$. These exponents also characterize the 
ordinary tricritical point in mean field theory. Thus the Lifshitz nature of 
the tricritical point does not affect these exponents within mean field 
theory. To see the difference between the Lifshitz and the ordinary 
tricritical 
point, we look at the behavior of deviations of the order parameter from 
uniformity. 
We find that $\Gamma_2^{(0)} 
({\bf q}\to 0)\sim t^{\gamma}$ with $\gamma=1$, again 
independent of the Lifshitz phenomena. However, if we rescale the wavevector 
${\bf q}$ by the correlation length 
$\xi$, we find that $\xi\sim t^{-\nu}$ with 
$\nu=1/4$ in contrast to ordinary critical behavior in which, with $g_0$ 
nonzero, $\nu$ would be 1/2. The Lifshitz behavior also is manifest in the 
value of the upper critical dimension, $d^*$, the smallest dimension at which 
mean field behavior is correct. For ordinary tricritical behavior, 
$d^*=3$, so that fluctuations are negligible. (They contribute logarithmic
corrections to mean field results.) For Lifshitz tricritical behavior, this is 
not so as $d^*$ is greater than 3. To determine this dimension, we employ a 
simple scaling argument applicable whenever fluctuations dominate the critical 
behavior. We rescale the 
order parameter $\eta$ by its bulk, uniform, value $\eta_0$ and 
the position vector ${\bf r}$ by the correlation length $\xi$ in Eq.(6.26). 
Again, 
$g_0$ and $b$ are set to zero so that the Lifshitz tricritical point is 
approached. Requiring that the coefficient of the 
Laplacian squared term be independent of the reduced temperature $t$, we 
obtain
$\eta_0 =\xi ^{(4-d)/2}\sim t^{(d-4)\nu/2}$, where $d$ is the  
dimension of space. However $\eta_0\sim t^{\beta}$ by the definition of the 
exponent $\beta$. Requiring the equality of these two exponents when the mean 
field values of $\beta=1/4$ and $\nu=1/4$ are inserted determines $d^*=6$.
Below this dimension, fluctuation effects are important, and the exponents 
differ from their classical values. These exponents are as yet unknown 
theoretically as well as experimentally, so that a measurement of any of them 
would be of considerable interest. One reason why they have not been 
calculated is that the usual epsilon expansion for this system fails at 
dimension d=4. This can be seen by considering a term of the form $\eta^p$ in 
the integrand of Eq (6.26). Again rescaling $\eta$ by $\eta_0$ and lengths by 
$\xi$ one finds the coefficient of such a term to be $\xi ^d\xi ^{p(4-d)/2}$.
At d=4, every term of the form $\eta^p$ has the same
coefficient $\xi ^d$. This means that one should use the whole series in
$\eta$ to perform reliable calculations, which would result in 
an infinite number
of renormalization group flow equations for the renormalized constants. 

Although the predictions of mean field theory fail at the Lifshitz tricritical 
point and within a region close to it, the predictions are valid outside this 
region, so that it is important to know its extent. This can be estimated with
the help of the Ginzburg criterion. We follow de Gennes$^{3)}$ and Joanny$^{5)}$ 
and assume that
fluctuations are unimportant if the 
fluctuations of the order parameter in a correlation
volume is much smaller than characteristic changes in 
the square of the order parameter itself, {\it i.e.} if
$$<(\delta\eta)^2>/\eta_0^2\propto 1/\Gamma^{(0)}_2(q\to 0)
\xi^d\eta_0^2\ll 1.\eqno(6.28)$$
Therefore, mean field theory is valid for temperatures $t_*$ such 
that
$$\displaystyle{{1\over {\Gamma_2^{(0)}(q\to 0)\xi ^d\eta_0^2}}\sim 
{{t_*^{d\nu-\gamma-2\beta}}\over {N^{d/2-1}}}\ll 1},\eqno(6.29)$$
where $\beta$, $\nu$ and $\gamma$ are the mean field exponents.
The N dependence follow from the fact that in polymers $\xi\sim N^{1/2}$ and
$\Gamma(0)\sim N$, where N is the number of monomers in the polymer. 
Here we
consider a system where all polymers have the same number of monomers.
For $d=3$, fluctuations can be ignored near the Lifshitz tricritical point 
provided that 
$$t_*\gg 1/N^{2/3}.\eqno(6.30)$$
Similarly, $t_*\gg 1/N$ for an ordinary critical point (see section III)
and $t_*\gg 1/N^{2/5}$
for a Lifshitz critical point.
In the limit N to infinity, mean field theory is valid in the 
entire critical region near the Lifshitz tricritical point. 
The influence of the Lifshitz on the surface tension of the coexisting
A-rich and B-rich phases (see Fig.23) is discussed in Appendix H.
The elastic properties of the interface between the A-rich and B-rich
phases can be find in Ref.[36].

\centerline{\bf VII. Rigid-flexible diblock copolymer system}

So far we have considered only the flexible systems. From Section II
we have learned that also the rigid polymers can be described using this
method. In this section we discuss the rigid-flexible diblock copolymer
system. The order parameter densities have been already discussed in
section II.D. They shall be used in the following subsections.
\vfill\eject

\centerline{\bf A. The scattering intensity}

The scattering matrix has the following form in the case of rigid 
flexible diblock copolymers$^{50)}$:
$$I_{ij}({\bf q})=\left(\Gamma_2^{(int)}({\bf q})+
\Gamma_2^{(0)}({\bf q})\right)_{ij}^{-1}.\eqno(7.1)$$
Here
$$\eqalign{&(\Gamma_2^{(0)}({\bf q}))^{-1}=I^{(0)}
({\bf q})=\cr &{1\over V}
\left\vert\matrix{
<\hat\phi_A^{(n)}\hat\phi_A^{(n)}>_0 & <\hat\phi_A^{(n)}\hat\phi_B^{(n)}>_0& 
<\hat\phi_A^{(n)}\hat Q_{\alpha\beta}^{(B)}>_0&<\hat\phi_A^{(n)}\hat Q_{\alpha\beta
}^{(A)}>_0\cr
<\hat\phi_B^{(n)}\hat\phi_A^{(n)}>_0 & <\hat\phi_B^{(n)}\hat\phi_B^{(n)}>_0& 
<\hat\phi_B^{(n)}\hat Q_{\alpha\beta}^{(B)}>_0&<\hat\phi_B^{(n)}\hat Q_{\alpha\beta
}^{(A)}>_0\cr
<\hat Q_{\alpha\beta}^{(B)}\hat\phi_A^{(n)}>_0 & 
<\hat Q_{\alpha\beta}^{(B)}\hat\phi_B^{(n)}>_0& 
<\hat Q_{\gamma\delta}^{(B)}\hat Q_{\alpha\beta}^{(B)}>_0&
<\hat Q_{\gamma\delta}^{(B)}\hat Q_{\alpha\beta
}^{(A)}>_0\cr
<\hat Q_{\alpha\beta}^{(A)}\hat\phi_A^{(n)}>_0 & 
<\hat Q_{\alpha\beta}^{(A)}\hat\phi_B^{(n)}>_0& 
<\hat Q_{\alpha\beta}^{(A)}\hat Q_{\gamma\delta}^{(B)}>_0&
<\hat Q_{\gamma\delta}^{(A)}\hat Q_{\alpha\beta
}^{(A)}>_0\cr}\right\vert .\cr}\eqno(7.2)$$
is the ideal part of the inverse vertex function and is 
solely determined by the conformations of a single copolymer molecule.
Matrix (7.2) is a symmetric twelve by twelve matrix. 
By the direct calculations of the
ideal averages, one may convince oneself that the averages involving the
nematic tensor for the flexible chain, $Q_{\alpha\beta}^{(A)}({\bf r})$, are
N$_0$ (N$_0$=N$_B$ or N$_A$) times smaller than the other averages. This is
understandable
since flexible chains consist of freely joined bonds, thus ordering one
bond does not affect the ordering of the other bonds. In contrast, in
the rigid rod, ordering one bond automatically orders the whole rigid 
B-part of the 
diblock copolymer. 
This allow us to neglect 
$Q_{\alpha\beta}^{(A)}$ in our calculations
and in the matrix(7.2). Thus $I^{(0)}$ is reduced to a seven by seven matrix.
Futhermore, by chosing the coordinate frame in which {\it 
the z axis is along the 
${\bf q}$ vector}, 
$I^{(0)}$ simplifies to the following block form:
$$I^{(0)}=\left\vert\matrix{{\Lambda}& 0\cr 0 & {\Delta}
\cr}\right\vert.\eqno(7.3)$$
Here 
$$\eqalign{&{\Lambda}=\cr &{1\over V}\left\vert\matrix{
<\hat\phi_A^{(n)}\hat\phi_A^{(n)}>_0 & <\hat\phi_A^{(n)}\hat\phi_B^{(n)}>_0& 
<\hat\phi_A^{(n)}\hat Q_{xx}^{(B)}>_0&<\hat\phi_A^{(n)}\hat Q_{yy
}^{(B)}>_0\cr
<\hat\phi_B^{(n)}\hat\phi_A^{(n)}>_0 & <\hat\phi_B^{(n)}\hat\phi_B^{(n)}>_0& 
<\hat\phi_B^{(n)}\hat Q_{xx}^{(B)}>_0&<\hat\phi_B^{(n)}\hat Q_{yy
}^{(B)}>_0\cr
<\hat Q_{xx}^{(B)}\hat\phi_A^{(n)}>_0 & 
<\hat Q_{xx}^{(B)}\hat\phi_B^{(n)}>_0& 
<\hat Q_{xx}^{(B)}\hat Q_{xx}^{(B)}>_0&
<\hat Q_{xx}^{(B)}\hat Q_{yy
}^{(B)}>_0\cr
<\hat Q_{yy}^{(B)}\hat\phi_A^{(n)}>_0 & 
<\hat Q_{yy}^{(B)}\hat\phi_B^{(n)}>_0& 
<\hat Q_{yy}^{(B)}\hat Q_{xx}^{(B)}>_0&
<\hat Q_{yy}^{(B)}\hat Q_{yy
}^{(B)}>_0\cr}\right\vert\cr} ,\eqno(7.4)$$
is a four by four matrix,  whereas $\Delta$ is a diagonal three by three 
matrix,
$${\Delta}={1\over V}
\left\vert\matrix{<\hat Q_{xy}^{(B)}\hat Q_{xy}^{(B)}>_0&0&0\cr
0&<\hat Q_{xz}^{(B)}\hat Q_{xz}^{(B)}>_0&0\cr
0&0&<\hat Q_{yz}^{(B)}\hat Q_{yz}^{(B)}>_0\cr}\right\vert.
\eqno(7.5)$$
Because the off diagonal elements of $\hat Q_{\alpha\beta}^{(B)}$
appears only in ${\Delta}$, they can be integrated out immediately
in the integrals (see Appendix A,B).

After the inversion of $I^{(0)}$, we will make use of the 
incompressibility
condition which, in Fourier space, reads
$$\phi_A^{(n)}({\bf q})+\phi_B^{(n)}({\bf q})=0, \eqno(7.6)$$
for ${\bf q}\ne 0$. After these algebraic manipulations, we end up with
the following form of the ideal and interaction vertex function: 
$$\Gamma_2^{(0)}=\left\vert\matrix{\Lambda_{11}^{-1}-2\Lambda_{12}^{-1}+
\Lambda_{22}^{-1}&\Lambda^{-1}_{23}-\Lambda_{13}^{-1}&\Lambda_{24}^{-1}
-\Lambda_{14}^{-1}\cr \Lambda_{23}^{-1}-\Lambda_{13}^{-1}&\Lambda_{33}^{-1}&
\Lambda_{34}^{-1}\cr \Lambda_{24}^{-1}-\Lambda_{14}^{-1}&\Lambda_{34}^{-1}&
\Lambda_{44}^{-1}\cr}\right\vert\eqno(7.7)$$
and
$$\displaystyle{\Gamma^{(int)}={\rho_0}
\left\vert\matrix{-2\chi&0&0\cr
0&-{4\over 3}v_{BB}&-{2\over 3}v_{BB}\cr
0&-{2\over 3}v_{BB}&-{4\over 3}v_{BB}\cr}\right\vert.}\eqno(7.8)$$

Here $\Lambda_{ij}^{-1}$ is the ij element of the inverse of $\Lambda$.
This matrix is also symmetric.

To summarize this section,
we have been able to reduce the set of twelve order parameters to the
relevant set of only three order parameters, namely the two previously
mentioned components of the nematic tensor i.e. $Q_{xx}^{(B)}({\bf q}), 
Q_{yy}^{(B)}({\bf q})$ and 
$\Psi({\bf q})$.

\centerline{\bf B. Stability limits of the isotropic phase}

At high temperatures the scattering matrix given by Eqs(7.1-8) is positive
definite thus the system is stable with respect
to compositional,
and nematic perturbations. 
However, as we lower the temperature
the scattering matrix ceases to be 
positive definite and the isotropic phase becomes unstable. 
We find the stability conditions by equating the determinant 
of the scattering 
matrix or its principal minors to zero.
The results of this stability analysis
are presented in Fig.25 as a plot of $\chi N$ 
versus the 
fraction of A monomers in the copolymer $f=N_A/N$.
If all phase transitions were continuous, 
these figures would be phase diagrams.
First we discuss the case in which there is no explicit interaction tending to 
orient the rigid rods, $v_{BB}=0$. This is shown in Fig. 25a.
One can easily see that the nematic order parameters can be integrated out 
in this case, and the stability condition is 
$$\displaystyle{
\Lambda_{11}+\Lambda_{22}+2\Lambda_{12}-
2\left
(\Lambda_{11}\Lambda_{22}-\Lambda_{12}^2\right)\rho_0\chi=0,}.\eqno(7.9)$$
The phase diagram encompasses in this case 
the disordered isotropic phase and 
some spatially ordered phases; for convenience we will call all of them
lamellar phases. The stability analysis within our random phase approximation
does not give us the
symmetry of these phases.
The stability condition against lamellar perturbations 
for the rigid-flexible system of the diblock copolymers (Eq(7.9))
has the same formal structure as that obtained in the
Random Phase Approximation 
by Leibler$^{22)}$ for the flexible-flexible 
diblock copolymer system. The difference between this condition and
the one obtained by Leibler is in the ideal averages which, in our case, 
reflect the rigid-flexible structure of the polymer
(see Appendix I).
Let us discuss the differences in the stability limits for these two systems.
First of all we note that
in the rigid-flexible copolymer
system the stability temperature is higher (or the interaction
Flory-Huggins $\chi$ parameter lower) in comparison to the systems of perfectly
flexible system
compare Fig.25a and Fig.12. 
For example below $\chi N=9,$ our system is stable for all f;
while the flexible-flexible diblock copolymer system
is only stable below $\chi N=10.5$. This reflects the reduction in 
entropy due to the rigidity of part of the polymer.
In addition, the mimimum of the curve 
shown in Fig.25a occurs at f=0.45 whereas in the flexible copolymers it occurs
at f=0.5. 
The results of this calculations find support in the computer 
simulation$^{94)}$.

For a system with nonzero
anisotropic interaction parameter, $v_{BB}$, a nematic phase
appears in addition to the lamellar 
and isotropic phases (Figs.25 bc)). 
The dashed line shown in Figs.25bc
denotes the stability limits against
nematic perturbations. It satisfies the following equation 
$$\displaystyle{{5\over{2N(1-f)^2}}-{v_{BB}\over {2}}=0.}\eqno(7.10)$$
Furthermore we observe that the 
stability limit against lamellar perturbations goes towards higher 
temperatures as the anisotropic interaction parameter increases, 
although the change is rather small, a few percent at most.
The black dot in the figures denotes the tricritical point at which
the isotropic 
phase is unstable against both nematic and lamellar perturbations 
at the same temperature. As the transitions in the system are actually first 
order, this tricritical point is preempted by a triple point at which all 
three phases coexist. The calculated tricritical point is an estimate of the 
location of the actual triple point.

\centerline{\bf C. The correlation functions}

We have calculated the following correlation functions
for {\bf q}=(0,0,$q_z$): 
$$\displaystyle{
\tilde G_{\Psi\Psi}(q_z)={1\over V}<\hat\Psi (q_z)\hat\Psi(-q_z)>;}
\eqno(7.11)$$
$$\displaystyle{
\tilde G_{\Psi Q}(q_z)={1\over V}<\hat\Psi (q_z)\hat Q_{zz}^{(B)}
(-q_z)>;}\eqno(7.12)$$
$$\displaystyle{
\tilde G_{QQ}(q_z)={1\over V}<\hat Q_{zz}^{(B)}(q_z)\hat Q_{zz}^{(B)}(-q_z)>;
}\eqno(7.13)$$
and the following Fourier transforms of these correlation functions:
$$\displaystyle{\eqalign{G_{\Psi\Psi}(z)=&
\int dx\int dy <\hat\Psi ({\bf r})\hat\Psi(0)>\cr
&={1\over {2\pi}}\int dq_z\tilde G_{\Psi\Psi}(q_z)\exp(iq_zz);\cr}}\eqno(7.14)$$
$$\displaystyle{\eqalign{G_{\Psi Q}(z)=&
\int dx\int dy <\hat\Psi ({\bf r})\hat Q_{zz}^{(B)}(0)>\cr
&={1\over {2\pi}}\int dq_z\tilde G_{\Psi Q}(q_z)\exp(iq_zz);\cr}}\eqno(7.15)$$
$$\displaystyle{\eqalign{G_{QQ}(z)=&
\int dx\int dy <\hat Q_{zz}^{(B)}({\bf r})\hat Q_{zz}^{(B)}(0)>\cr
&={1\over {2\pi}}\int dq_z\tilde G_{QQ}(q_z)\exp(iq_zz);\cr}}\eqno(7.16)$$
Here $Q_{zz}^{(B)}=-Q_{xx}^{(B)}-Q_{yy}^{(B)}$ measures the degree
of orientational ordering of the rigid part of the copolymer along the z-axis.
We note that as a direct consequence of the isotropy of the system
the scattering intensities (Eqs(7.11-13) can be rewritten in the 
following general
form:
$$\displaystyle{\tilde G_{\Psi\Psi}(q_z)=
{1\over V}<\hat\Psi ({\bf q}^*)\hat\Psi(-{\bf q}^*)>;
}\eqno(7.17)$$
$$\displaystyle{
\tilde G_{\Psi Q}(q_z)={1\over V}<\hat\Psi ({\bf q}^*){\bf \hat q}^*_{\alpha}
\hat Q_{\alpha\beta}^{(B)}({\bf q}^*){\bf\hat q}^*_{\beta}>;}\eqno(7.18)$$
$$\displaystyle{\tilde G_{QQ}(q_z)={1\over V}<{\bf\hat q}^*_{\gamma}
\hat Q_{\gamma\delta}^{(B)}({\bf q}^*){\bf\hat q}^*_{\delta}
{\bf \hat q}^*_{\alpha}
\hat Q_{\alpha\beta}^{(B)}({\bf q}^*){\bf \hat q}^*_{\beta}>.}
\eqno(7.19)$$
Here {\bf q}$^*$ is an arbitrary vector of length 
$\vert{\bf q}^*\vert$=$\vert q_z\vert$; ${\bf\hat q}^*$ is a unit vector
along {\bf q}$^*$. The summation over repeated indices is implied.

The calculations have been performed in the limit $q_z\rightarrow 0$,
$N_A,N_B\rightarrow\infty$, 
such that $q_zN_B=$const and $f=N_A/(N_A+N_B)=$const,
and for the following sets of parameters: (1) f=0.20, $\chi N=$19.51,
$v_{BB}/\chi=$0.4;
(2) f=0.20, $\chi N=$19.51, $v_{BB}/\chi=0$; (3) 
f=0.62, $\chi N=$11.35, $v_{BB}/\chi=$3.0; 
(4) f=0.62, $\chi N=$11.75, $v_{BB}/\chi=0$.
In all cases the parameters has been chosen such that the system is close to 
the stability limits which are: for case (1) $\chi N$=19.53; for case
(2) $\chi N$=19.56; for case (3) $\chi N=$11.40; and for
case (4) $\chi N=$11.79 (see Fig.25).

In Fig.26 $\tilde G_{\Psi\Psi}(q_z)$ and $G_{\Psi\Psi}(z)$ are shown for
f=0.20 and f=0.62.
In the case of a long rigid part, f=0.2 (Fig.(26a)),
there are two peaks at $q_1$ and $q_2$, 
whose positions are 
determined by the length of that part i.e. $q_1\approx 2\pi 
/((1-f)Nl)$ and
$q_2\approx 4\pi /((1-f)Nl)$. 
In the case of a short rigid part, f=0.62,
there is only one peak at $q_1$ (Fig(26c)).
The presence of the second harmonic in Fig(26a) and its absence in Fig(26c)
indicates that the domain boundaries are sharper in the
system containing longer rigid pieces. This is supported by 
the plots of the real space correlation functions. For f=0.62 (Fig.(26d))
this function is almost sinusoidal,
whereas it has additional structure for f=0.20 (Fig.(26b))
In both cases the length of the rigid part of a 
copolymer sets the length scale.
Finally by comparing the cases of $v_{BB}/\chi\not=0$ (solid line in Fig.(26))
and $v_{BB}/\chi=0$ (dashed line in Fig.(26))
we note that the general features of these functions are not affected
very much by the Maier-Saupe parameter $v_{BB}$. 

The nematic structure function, $\tilde G_{QQ}(q_z)$, and its Fourier 
transform (the nematic order parameter --- nematic order parameter
correlation function), $G_{QQ}(z)$, are shown in Fig(27) for f=0.62. 
The peak at $q_z=0$ 
reflects the tendency toward nematic order, while that at 
$q_z\approx 2\pi/fNl$ 
reflects the tendency to lamellar order. They appear in 
$G_{QQ}(z)$ as oscillations due to the nascent lamellar order
superimposed on the decaying nematic order. Not surprisingly, the Maier-Saupe 
interaction strongly affects the nematic signature of this 
correlation function as comparison with the case $v_{BB}/\chi=0$ clearly 
shows. For f=0.2 the tendency to nematic order is greatly
increased (because the system is closer to the isotropic-nematic
instability) 
so that $\tilde G_{Q Q}(q_z)$ is completely dominated by the peak at
zero wavevector, and the oscillations due to nascent lamellar order are
ignorable on the same scale. For this reason, we have not shown this function
or its Fourier transform.

In Fig.(28) the mixed correlation functions 
$\tilde G_{\Psi Q}(q_z)$ and $G_{\Psi Q}(z)$ are shown for
f=0.20 and f=0.62. 
Just as for the density-density correlation
function (Fig.(26)),
the structure of the density-nematic order parameter correlations is 
richer for f=0.20 than for f=0.62. Similarly,
these structures do not depend very strongly
on the Maier-Saupe parameter.
The most interesting observation to be made comes from a comparison of 
the functions $G_{\Psi \Psi}(z)$ and $G_{\Psi Q}(z)$ for f=0.62 
(Figs.(26d,28d)). We see that
the nematic order parameter in a given direction and the linear 
concentration of rigid rods in that direction are almost
exactly 
{\it anticorrelated}; {\it i.e.} 
large and positive $Q_{zz}^{(B)}(z)$ occurs in planes normal to z which
are 
rich in the {\it flexible} A-part of the  copolymer. 
At first sight, this result may appear to be
counter intuitive. One knows, for example, that in a mixture of rigid and 
flexible homopolymers, a higher density of the rigid polymers is found in the 
nematic phase. However we are concerned with the isotropic phase, and the
effect can be understood as arising from that isotropy.
As an example, let us assume that we are at a point in space in which 
the density of B monomers is greater than that of the A. For simplicity, we 
will assume 
that there are B monomers only out to a distance R, and 
A monomers from R to bR, with $b=(1-f)^{-1/3}$, 
so that the relative concentration of A 
monomers is f.
With such a density distribution, 
$$\eqalignno{G_{\Psi Q}(z)=({3V\over 4\pi n (bR)^3})\Bigl(\int_0^{
\sqrt{R^2-z^2}}&ds\  2\pi s
P_2({z\over \sqrt{z^2+s^2}})\cr
&-\int_{\sqrt{R^2-z^2}}^{\sqrt{(bR)^2-z^2}}ds\ 2\pi s
P_2({z\over \sqrt{z^2+s^2}})\Bigr).&(7.20)\cr}$$
Carrying out the integrations and setting $R=N_Bl/2=N(1-f)l/2,$ we obtain
$$\eqalignno{\rho_0 lG_{\Psi Q}(z)&=F(z/N_Bl),\cr
F(x)&=18x^2[{1\over 6}-\ln ({2x\over (1-f)^{1/3}})]
+{3\over 4}({1\over (1-f)^{2/3}}-2).&(7.21)\cr}$$
As long as the fraction of flexible monomers is not too large,
$f<1-2^{-3/2}\approx 0.65,$
this function is negative at zero argument, thus reproducing the anticorrelation of 
density and nematic fluctuations. Further, it has a maximum when the argument 
is approximately $0.36(1-f)^{1/3}$. In the above picture, we expect the 
spacing of spheres, and hence the period of the correlation function, 
 to be $N_Bl/(1-f)^{1/3}$. As the model 
correlation function is minimum at $z=0$, it should also 
be minimum at integer multiples of this period with maxima a
distance  $0.36(1-f)^{1/3}$ on either side. There are additional minima at 
half odd integer multiples of the period. For f=0.2 as in Fig. 28b, this 
model predicts 
minima at $z/lN_B=0,$ 0.54, 1.08, 1.62, 2.16, etc., and maxima at 
$z/lN_B=0.33,$ 0.75, 
1.41, 1.81, etc. These numbers are in satisfactory agreement 
with the figure. The double peak structure which this model, with its sharp 
density profile, produces requires at least the presence of a second harmonic 
to approximate it. As we have seen, the system with shorter rigid sections, 
f=0.62, does not exhibit a second harmonic in its structure function. Thus the 
above model does not describe the function of 
Fig. 28d very well at all except 
to reproduce its periodicity and the important 
negative value at the origin.
The above model should not be emphasized unduly, but it does serve to 
emphasize that the anticorrelation of density and nematic fluctuations that we 
have found is reasonable. 

The behavior of all correlation functions close to tricritical points 
is similar for all f$>$0.25, and 
in particular to the case f=0.62 presented here in 
detail. For f$\approx$0.25 the correlations functions begin to develop
the structure shown here for f=0.2. Finally we note that the locations
of the minima and maxima 
of the correlation functions are relatively independent 
on the Flory-Huggins
parameter. Changing it changes the heights of the peaks, but hardly affects
their positions. Thus the main features of the local domain structure
are determined by the geometry (rigidity and length)
of the copolymer molecule and excluded volume effects. In our calculation, the 
latter is modelled by assuming incompressibility of the 
system.
\vfill\eject

\centerline{\bf VIII. Rigid-flexible blends}

Here we shall show that within the approach
presented in this paper we can also
obtain the general
features of the nematic phase in
the system containing rigid and  flexible main-chain
nematogenic polymers.
The other systems like the  
mixtures of
two different rigid polymers, and a system of n-block copolymer 
consisting of n rigid and flexible
parts are described elsewhere$^{60)}$.
The Landau-Ginzburg
expansion of the free energy is derived, and we show that the
general feature of the isotropic nematic phase transition 
in the system of short flexible polymers and long rigid polymers
are qualitatively 
the same as obtained in the completely different treatment
of the density functional theory$^{95,96)}$ applied 
to infinitely long rods in the solvent. The latter model is
commonly known as the Onsager model$^{97)}$
The detailed study of the (extremely rich) phase diagrams produced in
the Landau-Ginzburg model and its variants is contained in
Ref[57,98-100].

In order to describe the system 
we shall use two order parameters:
the usual $\Psi$ parameter and a nematic order parameter
$Q_B=Q_{zz}^{(B)}$. The nematic order parameter for the
flexible part, $S_A=Q_{zz}^{(A)}$, can be neglected for
the reasons already stated in the previous section and in Ref[60].

The Landau-Ginzburg free energy for the rigid flexible
mixture (with the fraction of the 
monomers in the flexible polymers being
$\bar\phi$, can be easily calculated using the
approach presented in Appendix A and B. We find the 
following form of the free energy per monomer: 
$$ \Omega(Q_B,\Psi)=\Omega_I(\Psi)+\Omega_B(Q_B)
+\Omega_{C}(Q_B,\Psi).\eqno(8.1)$$
$\Omega_I$ is given by Eq(3.5);
$${\Omega_B(Q_B)=\left({5\over {2N_B(1-\bar\phi)}}-{{v_{BB}}\over {2}}\right) 
Q_B^2-
{{25}\over {21N_B(1-\bar\phi)^2}}
Q_B^3+{{425}\over {196N_B(1-\bar\phi)^3}}Q_B^4.}
 \eqno(8.2)$$
is the part of the free energy coming form the nematic
ordering of the rigid polymers;
$$\Omega_{C}
(Q_B,\Psi)={5\over {2N_B(1-\bar\phi)^2}}Q_B^2\Psi-{{50}\over {21N_B
(1-\bar\phi)^3
}}Q_B^3\Psi+{5\over {2N_B(1-\bar\phi)^3}}Q_B^2\Psi^2,\eqno(8.3)$$
is the term, representing the coupling between 
the order parameters in the mixture.

Given the free energy $\Omega(Q_B,\Psi)$, one easily finds
the conditions for thermal 
equilibrium of two or more 
coexisting phases. 
The condition
$${{\partial\Omega(Q_B,\Psi)}\over {\partial Q_B}}=0\eqno(8.4)$$
must be satisfied in the absence of external field in all
coexisting phases.  
Additionally the function
$$ G(Q_B,\Psi)\equiv 
\Omega(Q_B,\Psi)-{{\partial\Omega(Q_B,\Psi)}\over{\partial\Psi}}
\Psi.\eqno(8.5)$$
must be equal in all phases. 

For the two phase coexistance
the equations are solved for the three unknowns, 
$Q_B^{(1)}$, $Q_B^{(2)}$, and 
$\Psi$, i.e. 
the values of the order parameters in the two phases.
In the isotropic phase $Q_B=0$, but as we shall we we have
the cases when two nematic phases of different $Q_B$ can coexist.
Please note that
at coexistance we have in one phase $\Psi=0$. In the other
phase $\Psi$ is equal to the difference in the composition of the
coexisting phases.

We consider the mixture of 
flexible $A$ and rigid $B$ polymers in which the 
latter are much longer than the former,
$N_B\gg N_A$. At high temperatures one finds that the 
composition
difference between the B-rich nematic phase and A-rich isotropic
phase is extremely small, namely
$$\bar\phi_N-\bar\phi_I\sim {{N_A}\over {N_B}}\ll 1,\eqno(8.6)$$
where $\bar\phi_I$ ($\bar\phi_N$) is the concentration of A monomers in the 
isotropic (nematic) phase
at coexistence.
Neglecting terms of the order of $N_A/N_B$
or smaller, the two-phase coexistence region shrinks to a line given by
$$1\approx 0.21v^{IN}_{BB}(1-\bar\phi)N_B. \eqno(8.7)$$
where $v_{BB}^{IN}$ is the interaction parameter at the 
isotropic nematic phase transition. Please note that the parameter
is rescaled by $k_BT$.
The nematic order parameter jumps from zero in the isotropic phase to
$Q_B\approx 0.275(1-\bar\phi)$ in the nematic phase.
Here $\bar\phi$ is the average concentration of A monomers in the mixture at
N-I coexistence. Recalling that $Q_B$ is the nematic order parameter of the B 
monomers divided by the number of all monomers, we see from this result that 
the jump in nematic order parameter per B monomer is constant, independent of 
the 
concentration of the A component. These results are similar in 
spirit to the results obtained
by Warner and Flory$^{101)}$ and Onsager$^{97)}$ 
for the system of rods diluted
by a small particle solvent. 
If we assume that the monomers in the rigid polymer interact only via
excluded volume interactions we can write 
$v_{BB}\sim\rho_0v_0$  and $\rho_0(1-\bar\phi)=\rho_BN_B$, where
$\rho_B$ is the density of rigid polymers and $v_0$ 
is proportional to the excluded
volume of the two monomers. 
Since $v_0\sim l^2D_B$, where $l$ is the length of the monomer, $L_B=lN_B$
is the length of the rigid polymer and $D_B$ is the
thickness of the polymer we finally find the famous Onsager result
that in the limit of the rigid rod much larger than the solvent (either
monomeric or polymeric) and $L_B\gg D_B$,
the rod density at the isotropic-nematic
transition is inversly proportional to $L_B^2$ i.e.:
$$\rho_B L_B^2 D_B\sim 1\eqno(8.8)$$
Since the Onsager scaling is exact, Eqs(8.7-8) prove that the method 
applied here to rigid polymers is correct at the fundamental level.

We also find the phase separation inside the
nematic phase. This phase separation is goverened by the Flory Huggins
interaction parameter. The location of the critical point
is given in the limit of $N_B\gg N_A$ by (see Eq(3.6,3.7):
$$\chi_c^{mf}={1\over{2N_A}},\phantom{abc}\bar\phi_c^{mf}
=1-\sqrt{\left({{N_A}\over
{N_B}}\right)}\eqno(8.9)$$
As is evident from Eq.(8.9) the nematic ordering does not change 
the critical point temperature, which is the same as in the ordinary
polymer blend discussed in section III.

In this paper we have applied the Landau-Ginzburg model to
several systems: the polymer blends, diblock and multiblock 
copolymers,
rigid and flexible polymers and their mixtures.
The Landau-Ginzburg free energies and scattering intensities
were calculated from the
Edwards mesoscopic hamiltonian using the field theory methods.
We hope that this review will be helpful
for the future studies of polymer melts and blends.

\centerline{\bf Acknowledgements}

This work was supported by two grants from the Komitet Bada\'n Naukowych
and Fundacja Wsp\'o\l pracy Polsko-Niemieckiej.
R.H. acknowledges with appreciation the hospitality 
and support of the Max-Planck
Institute for Polymer Science in Mainz,
where the large part of this review has been prepared.
We would like to thank Anne Bohle, Michael Brereton, 
Wojciech G\'o\'zd\'z,
Andrzej Poniewierski, Michael Schick and Andrea Weyersberg
for fruitful collaboration and discussions. 

\vfill\eject

\centerline{\bf Appendix A: The cumulant expansion}

The method of cumulant expansion
is applied here to the binary mixture of
A,B homopolymers, but can be easily generalized to any other 
polymer system. 
The partition function, $Z[\phi^{(n_\gamma)}_\gamma]$ as a functional of 
prescribed Fourier transform of the concentration distribution,
$\phi^{(n_\gamma)}_\gamma({\bf q})$, ($\gamma= A,B$)is equal to
$$Z[\phi^{(n_\gamma)}_\gamma]=\exp{(-H[\phi^{(n_\gamma)}_\gamma]/k_BT)}
\bigr<\prod_{\gamma=A,B}\delta(\phi^{(n_\gamma)}_\gamma
({\bf q})-\hat \phi^{(n_\gamma)}_\gamma({\bf q})\Bigr>_0\eqno({\rm A.}1)$$
where the average over the conformations of the chains is
$$\displaystyle{<\cdots>_0=\prod_{\alpha=1}^
{n_A}\prod_{\beta=1}^{n_B}\int D{\bf r}^\alpha D{\bf r}^{\beta} 
\cdots W_A[{\bf r}^\alpha] W_B[{\bf r}^\beta]}.\eqno({\rm A.}2)$$
and the interaction hamiltonian, $H$, (see section II) is
$$\eqalign{
H[\phi^{(n_\gamma)}_\gamma]=&k_BT
\rho_0\int {{d{\bf q}}\over {(2\pi )^3}}\Bigr({1\over 2}w_{AA}
\vert\phi^{(n_A)}_A({\bf q})\vert^2\cr+&
{1\over 2}w_{BB}\vert\phi_B^{(n_B)}({\bf q})\vert^2+w_{AB}\phi_A^{(n_A)}({\bf q}) 
\phi_B^{(n_B)}(-{\bf q}))\Bigr).\cr}\eqno({\rm A.}3)$$

Now, we can introduce the integral representation of the delta function
in Eq.(A.2)  
$$\eqalign{\prod_{\gamma}&\delta(\phi^{(n_\gamma)}_\gamma({\bf q})-
\hat \phi^{(n_\gamma)}_\gamma({\bf q}))=\cr
&\prod_{\gamma}\int DJ_\gamma \exp\left(i\displaystyle{\int {{d{\bf q}}
\over{(2\pi )^3}}
(\phi^{(n_\gamma)}_\gamma({\bf q})-
\hat \phi^{(n_\gamma)}_\gamma({\bf q}))J_\gamma(-{\bf q})}\right)
\cr}\eqno({\rm A.}4)$$ 
and expand the exponential in $\hat \phi^{(n_\gamma)}_\gamma$.
 Eq.({\rm A.}1) now
reads
$$\displaystyle{\eqalign{
Z[\phi^{(n_\gamma)}_\gamma]=& \exp{(-H_I[\phi^{(n_\gamma)}_\gamma]/k_BT)}
\prod_{\gamma=A,B}
\Biggr(\int DJ_{\gamma}
\exp\left(i\int {{d{\bf q}}\over{(2\pi )^3}} \phi^{(n_\gamma)}_\gamma({\bf q})
J_\gamma(-{\bf q})
\right)\cr
\times & \sum_{n=0}^\infty {{(-i)^n}\over 
{n!}}
\int {{d{\bf q}_1}\over {(2\pi )^3}}
\cdots\int {{d{\bf q}_n}\over {(2\pi )^3}} S_{n}^{*(\gamma )}({\bf q}_1
\cdots{\bf q}_n)J_{\gamma}(-{\bf q}_1)\cdots J_{\gamma}(-{\bf q}_n)\Biggr),
\cr}}
\eqno({\rm A.}5)$$
where $\displaystyle{S^{*(\gamma )}_{n}}$ 
denotes the n-th moment 
$$\displaystyle{S^{*(\gamma )}_{n}({\bf q}_1\cdots{\bf q}_n)
=\Bigr<\hat \phi^{(n_\gamma)}_\gamma({\bf q}_1)\cdots\hat \phi^{(n_\gamma)}_\gamma({\bf q}_n)
\Bigr>_0}.\eqno({\rm A.}6)$$
It is obvious that ideal averages over A chain and over B chain 
can be performed independently, thus there are no mixed moments. 
The mixed moments are found only if the chains are connected inside a
molecule. This is the case of the diblock copolymer system.

The logarithm of the moments expansion is the cumulant expansion
$^{7,14)}$. 
Thus the second term in Eq.(A.5) can be rewritten as 
$$\displaystyle{\prod_{\gamma=A,B}\int DJ_{\gamma}
\exp\left(F^{(n_\gamma )}_\gamma [J_\gamma]+i
\int {{d{\bf q}}\over {(2\pi )^3}} \phi^{(n_\gamma)}_\gamma({\bf q})J_{\gamma}(-{\bf q})
\right)},\eqno({\rm A.}7)$$
where the cumulant expansion, $F^{(n_\gamma )}_\gamma[J_\gamma]$, which is 
a functional of 
the coupling fields, $J_\gamma$, is given by
$$\eqalign{
\displaystyle F^{(n_\gamma )}_\gamma
[J_\alpha]=&\sum_{n=1}^\infty{{(-i)^n}\over {n!}}
\int {{d{\bf q}_1}\over {(2\pi )^3}}\cdots\int 
{{d{\bf q}_n}\over {(2\pi )^3}} S^{(\gamma )}_{n}({\bf q}_1
\cdots{\bf q}_n)\cr &\delta ({\bf q}_1+\cdots +{\bf q}_n)
J_{\gamma}(-{\bf q}_1)\cdots J_{\gamma}(-{\bf q}_n).
\cr}\eqno({\rm A.}8)$$
It is easy to represent cumulants in terms of the moments$^{14,22}$
e.g.
$$\delta({\bf q}_1+{\bf q}_2)
S_2^{(A)}({\bf q}_1,{\bf q}_2)=S_2^{*(A)}({\bf q}_1,{\bf q}_2)-
S_1^{*(A)}({\bf q}_1)S_1^{*(A)}({\bf q}_2).$$
It is also easy to compute the moments and 
cumulants explicitely, e.g. 
$$\eqalign{&{1\over V}<\hat\phi^{(n_A)}_A ({\bf q})\hat\phi^{(n_A)}_A 
(-{\bf q})>_0=\cr &
{{n_A N_A}\over{\rho_0 V}}
\left({{1}\over {\rho_0N_A}}\right)
\left({{(1+a_A )}\over{(1-a_A)}}N_A
-{{2a_A}\over {(1-a_A )^2}}(1-a_A^{N_A})\right)\cr}$$
where $a_A =\sin{ql_A}/ql_A$ and $q\ne 0$. Expanding this
expression in $ql_A$ we find
$${1\over V}<\hat\phi^{(n_A)}_A ({\bf q})\hat\phi^{(n_A)}_A (-{\bf q})>_0=
{{n_A N_A}\over{\rho_0 V}}{{N_A}\over {\rho_0}}
g(x_A)+N_A{\rm O}(1/N_A),\eqno({\rm A}.9)$$
where $\bar\phi =n_AN_A/\rho_0V$ is the average concentration of A monomers in the
system,
$x_A=N_A (ql_A )^2/6$ and $g(x)=2(x+\exp (-x)-1)/x^2$
is the Debye function. Since we are interested in $1/\sqrt {N_\gamma}$
corrections we can neglect all higher order corrections
(e.g.$1/N_A$ corrections in Eq(A.9)).
To this point, no approximations have been made other than that which enabled 
us to express the 
conditional partition function 
in the form of Eq.(A.1) 

\centerline{\bf Appendix B: The saddle point approximation and RPA}  

Now the integral in expression (A.7) is calculated by the method of
the steepest decent, i.e. we approximate (A.7) as follows.
$$\eqalign{\int DJ_{\gamma} &
\exp\left(F^{(n_\gamma )}_\gamma[J_\gamma]+i
\int {{d{\bf q}}\over {(2\pi )^3}} 
\phi^{(n_\gamma)}_\gamma({\bf q})J_{\gamma}(-{\bf q})
\right)\approx\cr &
\exp\left(F^{(n_\gamma )}_\gamma[J^{*}_\gamma]+i
\int {{d{\bf q}}\over {(2\pi )^3}} \phi^{(n_\gamma)}_\gamma({\bf q})
J^*_{\gamma}(-{\bf q})
\right)\cr}\eqno({\rm B.}1)$$
Here, at the saddle point, the
coupling field, $J^*_{\gamma}({\bf q})$,
 satisfies the equation:
$$\displaystyle{{{\delta F^{(n_\gamma )}_\gamma
[J_\gamma]}\over {\delta J_\gamma(-{\bf q})}}
=-i\phi^{(n_\gamma)}_\gamma({\bf q})}.\eqno({\rm B.}2)$$
Thus, to within an uninteresting constant, the expression (A.7) is 
approximated by
$$\prod_{\gamma =A,B}\exp\left(
-\bar F^{(n_\gamma )}_\gamma[\Psi_\gamma]\right)=\prod_{\gamma =A,B}
\exp\left(F^{(n_\gamma )}_\gamma[J_\gamma^*]+i
\int {{d{\bf q}}\over{(2\pi )^3}}\phi^{(n_\gamma)}_\gamma({\bf q})J_{\gamma}^*(-{\bf q})
\right).\eqno({\rm B.}3)$$
Note that Eq.(B.3) together with Eq.(B.2) constitute 
the Lengendre transform.
  We have 
$\bar F^{(n_\gamma )}_\gamma[\Psi_\gamma ]$ :
$$\displaystyle\eqalign{\bar F^{(n_\gamma )}_\gamma[\Psi_\gamma]&
=\sum_{n=2}^\infty{{1}\over {n!}}
\int {{d{\bf q}_1}\over{(2\pi )^3}}\cdots\int {{d{\bf q}_n}
\over{(2\pi )^3}}\Gamma_{n}^{(\gamma)}({\bf q}_1
\cdots{\bf q}_n)\cr
\times&\delta ({\bf q}_1+\cdots +{\bf q}_n)\Psi_{\gamma}(-{\bf q}_1)      
\cdots \Psi_{\gamma}(-{\bf q}_n),\cr}\eqno({\rm B.}4)$$                   
with the coefficients, $\Gamma_n^{(\gamma )}$ which are simply            
expressed in terms of the cumulants$^{7,14,22}$ e.g.                      
$$\displaystyle{\Gamma_{2}^{(A)}({\bf q}_1,{\bf q}_2)=
1/S^{(A)}_{2}({\bf q}_1,{\bf q}_2)}.\eqno({\rm B.}5)$$
$$\displaystyle\Gamma_{3}^{(A)}({\bf q}_1,{\bf q}_2,{\bf q}_3)=
-{{S_3^{(A)}({\bf q}_1,{\bf q}_2,{\bf q}_3)}\over{
S^{(A)}_{2}({\bf q}_1,{\bf q}_2)S^{(A)}_{2}({\bf q}_2,{\bf q}_3)
S^{(A)}_{2}({\bf q}_1,{\bf q}_3)}}.\eqno({\rm B.}6)$$
The same formulas are obtained for B-chains.
The formulas for $\Gamma^{(\gamma)}_n$ functions for $n>3$ are
rather lengthy. Here we would like to give the abbreviated formulas
for $\Gamma_n^{(A)}, n=4,5$, needed in our later calculations. They are
as follows:
$$\Gamma_4^{(A)}=-\left(S_4^{(A)}-3\int S_3^{(A)}\Gamma_2^{(A)}S_3^{(A)}\right)
\left(\Gamma_2^{(A)}\right)^4,\eqno({\rm B}.7)$$
$$\Gamma_5^{(A)}=-\left(S_5^{(A)}+15\int S_3^{(A)}\Gamma_2^{(A)}\int S_3^{(A)}
\Gamma_2^{(A)}S_3^{(A)}-10\int S_4^{(A)}\Gamma_2^{(A)} S_3^{(A)}\right)
\left(\Gamma_2^{(A)}\right)^5,\eqno({\rm B}.8)$$
where the proper symmetrization and the integration over dummy variables
are assumed.

We note that Eq.(A.1) together with  
Eq.(B.4) constitute
the basis of the 
random phase approximation (RPA)$^{2,22)}$ and the RPA vertex functions
$\Gamma^{(0)}_n$ are easily obtained by summing the interaction term form
the hamiltonian and the term involving $\Gamma^{(A)}_n$ and $\Gamma^{(B)}_n$. 
For example we have
$$\Gamma^{(0)}_2=\Gamma_n^{(A)}+\Gamma_2^{(B)}-2\chi\rho_0
\eqno({\rm B.}9)$$
and
$$\Gamma^{(0)}_3=\Gamma_3^{(A)}-\Gamma_3^{(B)}.\eqno({\rm B}.10)$$ 
In the limit of
$q\to 0$ we find 
$$\Gamma_3^{(A)}(0,0,0)=-{1\over{\bar\phi}}
\Gamma_2^{(A)}(0,0)\eqno({\rm B.}11)$$
(analogously for B with $\bar\phi$ changed to $1-\bar\phi$). The 
right hand side 
of Eq(B.11) is given by the inverse of Eq(A.9) (see Eq(3.9)).
\vfill\eject
\centerline{\bf Appendix C: One-loop equations for the vertex functions}

Fig.7 and Fig.8 show the one-loop 
diagrams contributing to $\Gamma_2$ and $\Gamma_3$,
respectively. Note that $\Gamma_4^{(0)}$ and $\Gamma_3^{(0)}$ contribute
to $\Gamma_2$ (Fig.7), while $\Gamma_3^{(0)}$ and $\Gamma_5^{(0)}$
contribute to $\Gamma_3$(Fig.8). The solid lines represent $1/\Gamma_2^{(0)}$
while the higher order vertex functions 
are represented
by points in the diagrams. The n-body vertex function has n lines 
emanating from the point representing this function.
The dashed lines
are not parts of the diagrams
but are included only for clarity (according to Ref.[7]).
The final approximation (Hartree) in the framework of the loop expansion
is to make the equations self consistent. Here it corresponds 
to the change of
$\Gamma_2^{(0)}$ and $\Gamma_3^{(0)}$ appearing in the integrals,
represented by the diagrams in Figs.(7,8), to $\Gamma_2$ and $\Gamma_3$.
The final self-consistent one-loop equations for the 
two-body and three-body vertex functions are as follows:
$$\eqalign{&\Gamma_2({\bf q},-{\bf q})=\Gamma_2^{(0)}({\bf q},-{\bf q})
+{1\over 2}\int{{d{\bf k}}\over{(2\pi )^3}}
{{\Gamma_4^{(0)}({\bf q},-{\bf q},{\bf k},-{\bf k})}\over {\Gamma_2
({\bf k},-{\bf k})}}\cr &-{1\over 2}\int{{d{\bf k}}\over{(2\pi )^3}}
{{\left(\Gamma_3({\bf q},{\bf k},-{\bf q}-{\bf k})\right)^2}\over {\Gamma_2
({\bf k},-{\bf k})\Gamma_2({\bf q}+{\bf k},-{\bf q}-{\bf k})}},\cr}\eqno({\rm C.}1
)$$
$$\eqalign{&\Gamma_3({\bf q},{\bf p},-{\bf q}-{\bf p})=\Gamma_3^{(0)}
({\bf q},{\bf p},-{\bf q}-{\bf p})\cr &+\int{{d{\bf k}}\over{(2\pi )^3}}
{{\Gamma_3({\bf q},{\bf k},-{\bf q}-{\bf k})
\Gamma_3({\bf p},{\bf k},-{\bf p}-{\bf k})
\Gamma_3(-{\bf q}-{\bf p},{\bf p}+{\bf k},{\bf q}-{\bf k})}\over {\Gamma_2
({\bf k},-{\bf k})\Gamma_2({\bf q}+{\bf k},-{\bf q}-{\bf k})
\Gamma_2({\bf p}-{\bf k},-{\bf p}+{\bf k})}}\cr &+
{1\over 2}\int{{d{\bf k}}\over{(2\pi )^3}}{{\Gamma_5^{(0)}
({\bf k},-{\bf k},{\bf q},{\bf p},-{\bf q}-{\bf p})}\over
{\Gamma_2({\bf k},-{\bf k})}}.\cr}\eqno({\rm C}.2)$$
Before proceeding, we first want to determine the order (in 
terms of $N_\gamma$, $\gamma
=A,B$) of successive terms in Eqs.(C.1,2). For simplicity, we consider for a moment
the case of $N_A=N_B=N$ and $l_A=l_B=l$. For equal length of A and B
polymers the critical concentration is $\bar\phi_c=1/2$ and at the critical
point only the first and the second terms in Eq.(C.3) are nonzero.
By comparing these two terms we find the following
relation at the critical point:
$${1\over {\vert\Gamma_2^{(0)}(0,0)\vert}}\int{{d{\bf k}}\over{(2\pi )^3}}
{{\Gamma_4^{(0)}({\bf k},-{\bf k},0,0)}\over{\Gamma_2({\bf k},-{\bf k})}}
\sim {l\over{\Lambda}}+N{{l^3}\over{\Lambda^3}},\eqno({\rm C}.3)$$
where $2\pi /\Lambda$ is the upper cutoff in the integral. As we can see
the choice of the
cutoff is crucial for determining the relative magnitude of the
mean-field value of $\Gamma_2$ (that is $\Gamma_2^{(0)}$) and the
first (one-loop) correction to it. There are no ready recipes for
the choice of the cutoff. Here we take it proportional
to the radius of gyration i.e. to $\sqrt{N}$. The discussion  
of the cutoff is
given in Appendix D.   

In the solution of Eq.(C.1,2)  we
use 
the following approximation for 
$\Gamma_2^{(A)}$ (similarly for $\Gamma_2^{(B)}$)
(see Appendix A and B):
$$\Gamma_2^{(A)}({\bf q},-{\bf q})={{1+{R_A^2q^2/2}}\over{\bar\phi N_A}}
.\eqno({\rm C}.4)$$
The coefficient  (Eq.C4) in front of $q^2$ is obtained
from the expansion of the expression
(A.9) for large $ql_A$.
 Finally we observe that
$$\Gamma_3^{(A)}({\bf q},-{\bf q},0)=
-\Gamma_2^{(A)}({\bf q},-{\bf q})/\bar\phi,
\eqno({\rm C}.5)$$
$$\Gamma_4^{(A)}({\bf q},-{\bf q},0,0)=2\Gamma_2^{(A)}({\bf q},-{\bf q})
/\bar\phi^2,\eqno({\rm C.}6)$$
$$\Gamma_5^{(A)}({\bf q},-{\bf q},0,0,0)=-6\Gamma_2^{(A)}({\bf q},-{\bf q})
/\bar\phi^3.\eqno({\rm C}.7)$$
We have analogous equations for B chains with $\bar\phi$ changed to
$1-\bar\phi$. 

\centerline{\bf Appendix D: The upper wavevector cutoff in polymer blends}

As we have seen in section III the fluctuation corrections to the
critical temperature strongly depend on the upper wavevector cutoff.
Although it is ubiquitous in statistical mechanics
there are no ready recipes for
the choice of the cutoff. In low molecular mass liquids we usually have one 
natural length scale, which corresponds to the size of a molecule and
 the cutoff
is usually made proportional to this length scale. In the polymer
mixtures the problem is more complicated since we have three different length
scales: the total length of a polymer molecule, $Nl$,
the size of the region occupied by a polymer molecule
(proportional to the radius of gyration), $\sim\sqrt{N}l$, and finally,
the microscopic length scale $l$, which is determined by 
the size of a single monomer. We believe that $\Lambda$ should be proportional to the
radius of gyration, that is to $\sqrt{N}$. 
At high temperatures, where the interactions between the monomers
are irrelevant,
the monomer-monomer correlation function decays
exponentially with the characteristic correlation length which is
proportional to the radius of gyration. 
Moreover, in all our calculations the specific
structure of monomers has not been taken into account.  
The microscopic length scale is
irrelevant here. Finally, if we chose the microscopic length scale as 
the cutoff
the fluctuation corrections would survive in the limit of
$N\to\infty$ and thus RPA would not be the correct description in this
limit. We believe this is not the case. 
We note that since the size of the polymer molecule in the blend is 
roughly
proportional to $\sqrt{N}$; this choice is also in accordance with
the prescription known from low molecular mass systems where the cutoff is 
made proportional to the size of a molecule.
 In the case of assymetric mixture we postulate
that the cutoff should be equal to $2\pi/(\Lambda(R_A,R_B))$, where
$\Lambda$ is the symmetric function of the two radiuses of
gyrations for A and B chains.
{\it Finally, we assume the scaling form for $\Lambda$ }:
$$\Lambda(R_A,R_B)=R_Af(x),\eqno({\rm D}.1)$$
where $x=R_B/R_A$. From the symmetry properties of
$\Lambda$, we have the following equation for $f(x)$:
$${1\over x}f(x)=f\left({1\over x}\right).\eqno({\rm D}.2)$$
The general solution of this functional equation is
$$f(x)=\sqrt{C_1{{1+x^3}\over{1+x}}+C_2x}.\eqno({\rm D}.3)$$
Here $C_1$ and $C_2$ are two constants which depend on the 
microscopic details of the system.
We cannot determine them from the present
theory, thus, in the interpretation of experimental
results, they should be the fitting parameters. 

In the simplest approximation we can set $R_A\sim\sqrt{N_A}l_A$. In general
in the polymer system this radius changes with temperature
and concentration and so it would be
desirable here to determine it self consistently from the
theory.
The equation for the
radius of gyration, $R$,
offers a simple way for the self consistent determination of the
cutoff. if we set it equal to $2\pi C/R$ where $C$ is some numeric
constant. Thus apart from this constant the cutoff would be determined from the
theory. In section III D the equations for the radius of gyration are
shown to involve the collective structure factor (proportional to
the scattering intensity Eq(3.9)).
 In this way the equation for the collective structure factor
and the equation for the radius of gyration 
are mutually coupled. 
 
\centerline{\bf Appendix E}

Here we present the results used in section IIIC for 
the calculation of the radius of gyration.
Here we shall follow the equations given in Ref[20].
For the fourth order correlation function we have:
$$\eqalign{&<\hat\phi^{(1)}_A({\bf q})\hat\phi_A^{(1)}(-{\bf q})
\hat\phi^{(1)}_A({\bf k})\hat\phi_A^{(1)}(-{\bf k})>_0=
8g_1(q,0,k)+2(g_1(q,{\bf q}+{\bf k},q)\cr&+g_1(q,{\bf q}-{\bf k},q)+
g_1(k,{\bf q}+{\bf k},k)+g_1(k,{\bf q}-{\bf k},k))\cr &+
4(g_1(q,{\bf q}+{\bf k},k)+g_1(q,{\bf q}-{\bf k},k)).\cr}\eqno({\rm E}.1)$$
Here 
$$\eqalign{g_1&(q_1,q_2,q_3)={{N_A}\over{x_1x_2x_3}}+{{p(x_1)}\over
{x_1(x_1-x_2)(x_1-x_3)}}\cr+&
{{p(x_2)}\over{x_2(x_2-x_3)(x_2-x_1)}}+{{p(x_3)}\over{x_3(x_3-x_2)(x_3-x_1)}}
,\cr}\eqno({\rm E}.2)$$
where $x_i=q_i^2l^2/6$ and $p(x)=(\exp{(-xN_A)}-1)/x$. For the
two point correlation function we find
$$<\hat\phi^{(1)}_A({\bf q})\hat\phi_A^{(1)}(-{\bf q})>_0=
2(N_A+p(x))/x,\eqno({\rm E}.3)$$
and $x=q^2l^2/6$.

In order to obtain the radius of gyration one has to differentiate
$$\eqalign{h({\bf k},{\bf q})&=<\hat\phi^{(1)}_A({\bf q})\hat\phi_A^{(1)}(-{\bf q})
\hat\phi^{(1)}_A({\bf k})\hat\phi_A^{(1)}(-{\bf k})>_0\cr &
-<\hat\phi^{(1)}_A({\bf q})\hat\phi_A^{(1)}(-{\bf q})>_0
<\hat\phi^{(1)}_A({\bf k})\hat\phi_A^{(1)}(-{\bf k})>_0\cr}
\eqno({\rm E}.4)$$
twice with respect to q and finally take the limit of
$q\to 0$. We find
$$\eqalign{h''({\bf k},0)=&f({\bf k})=4N_A^5\cos^2\theta(-120+120\exp{(y)}
-96y-24\exp{(y)}y-24y^2\cr-&12\exp{(y)}y^2+4\exp{(y)}y^3+y^4)/(3\exp{(y)}y^5)
,\cr}\eqno({\rm E}.5)$$
where $y=N_Ak^2l^2/6$ and $\theta$ is the angle between the
${\bf q}$ and ${\bf k}$ vector.

\centerline{\bf Appendix F: One-loop equation for $\Gamma_2$ in
the lamellar phase}

In the lamellar phase of the
diblock copolymer system the field $\Psi$ is approximated 
by the single harmonic:
$$\Psi=Acos(q_lz)\eqno({\rm F.}1)$$
In general $q_l$, which describes the periodicity of the phase, is
different from the $q^*$, which marks the divergence of the structure
factor. In practice they are very close.
The vertex function in this case, $\Gamma_2({\bf q},A)$ is
the function of the amplitude $A$ and the wavevector ${\bf q}$.
In the disordered phase A=0, and $\Gamma_2$ depends only on the
modulus of ${\bf q}$. The equation for the vertex function for the 
symmetric case ($f=1/2$) is
as follows$^{20)}$:
$$\eqalign{&\Gamma_2({\bf q},A)=\Gamma_2^{(0)}(\vert{\bf q}\vert,A=0)
+{1\over 2}\int{{d{\bf k}}\over{(2\pi )^3}}
\Gamma_4^{(0)}({\bf q},-{\bf q},{\bf k},-{\bf k})\Gamma_2^{-1}
({\bf k},A)\cr &+A^2\Gamma_4^{(0)}({\bf q},{\bf -q},q_l,-q_l)
\cr}\eqno({\rm F.}2
)$$
Please note that the function, $\Gamma_2$ is not isotropic in the lamellar
phase, because a particular direction (z in this case (Eq(F.1))) 
has been selected. The dependence of $\Gamma_4^{(0)}$ on
angles is rather weak.  
\vfill\eject

\centerline{\bf Appendix G}

The coefficent in the Landau-Ginzburg free energy Eqs(6.26-27) are as
follows:
$$\eqalignno{a&={\gamma_1}\cr
             b&={2\gamma_2\gamma_3-\gamma_4^2\over4\gamma_3}\cr
             c&={4\gamma_3^3\gamma_5-3\gamma_3^2\gamma_4\gamma_8
+3\gamma_3\gamma_4^2\gamma_6-\gamma_4^3\gamma_7\over12\gamma_3^3}.&({\rm
G.}1 )\cr}$$
where  
$$\eqalignno{\gamma_1&={1\over(1-\phi)}-{\chi N\over 2},\cr
             \gamma_2&={1\over3(1-\phi)^3},\cr
             \gamma_3&={1\over\phi(1-\phi)},\cr
             \gamma_4&={1\over(1-\phi)^2},\cr
             \gamma_5&={1\over5(1-\phi)^5},\cr
             \gamma_6&={1\over(1-\phi)^3},\cr
             \gamma_7&=-{(1-2\phi)\over2(1-\phi)^2\phi^2},\cr
             \gamma_8&={1\over(1-\phi)^4}.&({\rm G.}2)\cr}$$
Note that with the exception of $\gamma_1$, all coefficients are determined 
solely by the entropy of mixing.

 The consolute line is given by the vanishing of the coefficient $a$ in Eq. 
 or, from Eqs. (G.1) and (G.2),
$$\chi N={2\over {(1-\phi)}}.\eqno({\rm G}.3)$$
A tricritical point arises when the coefficients $a$ and $b$ vanish 
simultaneously, which occurs at the point
$$\eqalignno{\chi_{tri} N&=6,\cr
             \phi_{tri}&=2/3.&({\rm G}.4)\cr}$$
 We have established, therefore, that this point, the point at which the 
disorder line and Lifshitz lines all meet, is a tricritical point. To 
establish that it is a Lifshitz tricritical point we have to check $g_0$.
Here we have
$$g_0={1\over 3}\Bigr ({1\over {(1-\phi )}}-{{(\chi N)^2\phi }\over 8}\Bigr)
{{Nl^2}\over 6},\eqno({\rm G.}5)$$
$$h_0={1\over 36}\Bigr ({1\over {(1-\phi )}}+{{9(\chi N)^2\phi }\over 16}
-{{(\chi N)^3\phi^2}\over 8}\Bigr ){{N^2l^4}\over{36}}.\eqno({\rm G.}6)$$
We immediately verify that at the tricritical point 
$N\chi_{tri}=6$, $\phi_{tri}=2/3$ at 
which the coefficients $a(\chi,\phi)$ and $b(\chi,\phi)$ both vanish, the 
coefficient $g_0(\chi,\phi)$ also vanishes. Therefore the point is a Lifshitz 
tricritical point as stated (see also Appendix H).

\centerline{\bf Appendix H: Influence of Lifshitz Tricritical Point}

 At values of $\chi$ larger than those given by the consolute line 
(in section VI), the system separates into A-rich and B-rich phases
(Fig.23). We shall calculate 
approximately the surface energy between these two phases to illustrate the 
effect of the proximity of the Lifshitz tricritical point. 
The two coexisting phases are characterized by uniform order parameters of
amplitude $\pm\eta_0$, where this amplitude is obtained by minimizing 
the free 
energy of Eq. (6.26) and requiring a uniform solution. One obtains
$$\eta_0^2={-b+\sqrt{b^2-3ac}\over 3c}.\eqno({\rm H.}1)$$
To determine the interfacial profile $\eta(z)$ between these phases, we 
require
a non-uniform solution 
of the Euler-Lagrange equation which expresses the minimization of
the free energy Eq. (6.28). The solution must also satisfy 
the boundary 
conditions
$$\eta (\pm\infty )=\pm\eta _0.\eqno({\rm H.}2)$$
The Euler-Lagrange equation is of fourth order and  nonlinear, and we are 
unable to solve it.
In order to obtain an analytic form for the surface free energy, we 
approximate the bulk part of the free energy by a double parabola with the 
same curvature at the minima; 
$$ 
{1\over 2}(-|a|\eta^2({\bf r})+b\eta^4({\bf r})+c\eta^6({\bf r}))
\approx\cases{{2|a|y(s)}(\eta({\bf r}) -\eta_0)^2  &if $\eta >0$,\cr
{2|a|y(s)}(\eta({\bf r}) +\eta_0)^2
&if $\eta <0$,\cr}\eqno({\rm H.}3)$$
where 
$$y(s)\equiv[(1+s^2)^{1/2}-s](1+s^2)^{1/2},\eqno ({\rm H.}4)$$
and 
$$s\equiv b/(3|a|c)^{1/2}.\eqno({\rm H.}5)$$
The parameter $s$ is a scaling field, 
a dimensionless measure of the deviation, $b$, of the system from tricritical 
behavior measured with respect to $a^{1/2}$, where $a$ is proportional to 
$t\equiv(T-T_{con})/T_{con}$, the deviation of the temperature from 
the consolute temperature.
With this approximation in the free energy, the Euler-Lagrange equation is 
piecewise linear 
$$h_0{d^4\eta\over dz^4}-g_0{d^2\eta\over dz^2}+{4|a|y(s)}
(\eta\pm\eta_0)=0.\eqno({\rm H.}6)$$
The form of the solution depends upon the ratio $y(s)/p^2$
where $p$ is a second scaling field
$$p\equiv g_0/(4|a|h_0)^{1/2},\eqno({\rm H.}7)$$
a dimensionless measure of the amplitude of the square gradient term in the 
free energy. It is a measure of the deviation of the system from Lifshitz 
behavior. The ratio of $y/p^2$ is easy to interpret; the scaling field 
$p$ for fixed $g_0$ diverges as the consolute line is approached. Similarly, 
the scaling field $s$ for fixed $b$ also diverges as 
the consolute 
line is approached. However, the function $y(s)$, only varies by a factor of 
two as $s$ varies from zero to infinity. Thus $y/p^2$ will always be smaller 
than unity sufficiently close to the consolute line; how close in terms of 
temperature depends on the 
amplitude $g_0$, that is, how close the system is to 
Lifshitz behavior.

In general the form of the profile is
$$\eta (z)=\cases {-\eta _0+K_1\exp {(z/\xi _+)}+K_2
\exp {(z/\xi _-)}, & if $z<0$;\cr 
\eta _0-K_1\exp {(-z/\xi _+)}-K_2\exp {(-z/\xi _-)}, & if $z>0$.\cr}
\eqno({\rm H.}8)$$
with
$$K_1={{\eta _0(\xi _+/\xi _-)^2}\over {(\xi _+/\xi _-)^2-1}},
\eqno({\rm H.}9)$$
$$K_2=-{{\eta _0}\over {(\xi _+/\xi _-)^2-1}},
\eqno({\rm H.}10)$$ 
and two decay length in the system
$$\xi_{\pm}^{-2}=2\left({{|a|}\over h_0}\right)^{1/2}p\bigl(1\pm(1-{y\over 
p^2})^{1/2}\bigl).\eqno({\rm H.}11)$$
As the consolute line is approached, the correlation length 
$\xi_{-}$ diverges as $t^{-1/2}$, 
while $\xi _+$ is finite unless the system approaches the Lifshitz 
tricritical (or critical) point.
It is easy to see that for $y/p^2<1$, $\xi_{\pm}$ are real and
the profiles are monotonically decaying functions
at large distances from the interface.
For $y/p^2>1$, that is, sufficiently close to the Lifshitz tricritical point,
$\xi_{\pm}$ are complex and the profiles are no longer monotonically 
decaying, but are damped oscillatory 
functions. Using Eq({\rm H.}8-{\rm H.}11) this oscillatory profile 
can be written in the following form: 
$$\eta(z)=\cases{-\eta_0+\exp{(z/\xi)}(\eta_0\cos({2\pi z/\lambda})+K_3
\sin({2\pi z/\lambda}))& if $z<0$,\cr
\eta_0+\exp{(-z/\xi)}(-\eta_0\cos({2\pi z/\lambda})+K_3
\sin({2\pi z/\lambda}))& if $z>0$,\cr}\eqno({\rm H.}12)$$
where 
$$K_3={1\over {\sqrt{y/p^2-1}}},\eqno({\rm H.}13)$$
$$\xi^{-2}=\left({{|a|}\over {h_0}}\right)^{1/2}p\left(1+{y^{1/2}\over p}
\right),
\eqno({\rm H.}14)$$
and$$({2\pi\over\lambda})^2=\left({{|a|}\over{h_0}}\right
)^{1/2}{p}\left(1-{y^{1/2}\over p}\right),
\eqno({\rm H.}15)$$
Note that this profile has the similar 
form as the correlation function of 
the disordered phase on the copolymer-rich side of the disorder line
(section VI B). 
We note here that the non-monotonic 
profiles are encountered in homopolymer-rich 
phases sufficiently close to the Lifshitz tricritical point. The correlation 
length $\xi$ diverges as the Lifshitz tricritical point is approached as 
$t^{-1/4}$, as expected from Sec.VI C.

The interfacial tension is obtained by inserting these profiles
into Eq. (6.26) and dividing by the area of the planar interface.
The result can be written in the scaling form
$$\sigma(t,b,g_0)=\left({{|a|^5h_0}\over {9c^2}}\right)^{1/4}{2\rho_0k_BT 
\over N}\Sigma(s,p),\eqno({\rm H.}16)$$
where
$$\Sigma(s,p)=\cases{\displaystyle{{{y^{3/2}}\over{(1+s^2)^{1/2}}}{(2p+y^{1/2})
\over
[p-(p^2-y)^{1/2}]^{1/2}+[p+(p^2-y)^{1/2}]^{1/2}}},& if $y\le p^2$\cr
\displaystyle{{y^{3/2}\over (1+s^2)^{1/2}}{(2p+y^{1/2})\over(p+y^{1/2})^{1/2}}}
& if 
$y\ge p^2$.\cr}\eqno({\rm H.}17)$$
The surface tension is proportional to $N^{-1/2}$ as $h_0$ is proportional to 
$N^2$.
This expression for the surface tension has the 
following properties; on approaching the Lifshitz tricritical point, 
$t\rightarrow 0$, p and s constant,
$$\sigma(t,s,p)\sim t^{5/4};\eqno({\rm H.}18)$$
on approaching an ordinary tricritical point, $t\rightarrow 0$, s constant, 
$p\sim t^{-1/2}$,
$$\sigma(t,s,p)\sim t;\eqno({\rm H.}19)$$
on approaching a Lifshitz critical point, $t\rightarrow 0$, $s\sim t^{-1/2}$, 
p constant,
$$\sigma(t,s,p)\sim t^{7/4};\eqno({\rm H.}20)$$
on approaching an ordinary critical point, $t\rightarrow 0$, s/p constant, 
$s\sim t^{-1/2}$,
$$\sigma(t,s,p)\sim t^{3/2}.\eqno({\rm H.}21)$$
All of these behaviors are the ones expected in mean field theory. The 
crossover between these behaviors is governed by the two scaling fields
$s$ and $p$ of Eqs. ({\rm H.}5) and ({\rm H.}7).

Sufficiently near a transition so that fluctuations become important, 
the surface tension will still have a 
scaling form
$$\sigma(t,b,g_0)=t^{\mu}\Sigma\left({b\over t^{1/\phi_t}},{g_0\over 
t^{1/\phi_g}}\right),\eqno({\rm H.}22)$$
but the form of the scaling function $\Sigma$ will differ from that given 
above, and  the 
critical exponent $\mu$ and the crossover exponents $\phi_t$ and 
$\phi_g$ will differ from their mean-field values 5/4, 2, and 2, respectively.
As noted earlier, there are no calculations as to the values these exponents 
should take.

The presence of the Lifshitz tricritical point also affects the results of 
scattering measurements as seen in the structure function. At long 
wavelengths, and sufficiently close to the consolute line, the form of a 
general structure factor is
$$I(q)= (a+g_0q^2+Kq^4)^{-1},\eqno({\rm H.}23)$$
where $K$, a function of the copolymer concentration, will depend on the 
particular structure function. On approaching any point on the consolute line 
except the Lifshitz tricritical point, it is convenient to rewrite this 
expression in the form
$$\eqalignno{I(q)&= a^{-1}(1+{g_0\over a}q^2+{Ka\over g_0^2}{g_0^2\over 
a^2}q^4)^{-1}\cr
&=a^{-1}(1+(q\xi)^2+K^{\prime}p^{-2}(q\xi)^4)^{-1},&({\rm H.}24)\cr}$$
where, to within numerical factors, $\xi$ is the ordinary correlation length 
$\xi_{-}$ of Eq. ({\rm H.}9). 
As the scaling field $p$ diverges as the consolute 
line is approached, the term in $q^4$ becomes unimportant and the structure 
function can be written in the scaling form
$$\eqalignno{I(t,q)&=ts(q\xi)\cr
             s(x)&=(1+x^2)^{-1},\qquad p\rightarrow\infty&({\rm H.}25)\cr}$$    
that is, the usual Lorentzian form.
On the other hand, if one approaches the Lifshitz tricritical point so that
the scaling field $p$ remains finite, it is more convenient to write Eq. 
({\rm H.}23) as
$$\eqalignno{I(q)&=a^{-1}(1+{g_0\over (aK)^{1/2}}({K\over a}^{1/4}q)^2+({K\over 
a}^{1/4}q)^4)^{-1},\cr
&=a^{-1}(1+K^{\prime\prime}p(\xi q)^2+(\xi q)^4)^{-1},&({\rm H.}26)\cr}$$
where $K^{\prime\prime}$ is an unimportant function of the copolymer 
concentration and $\xi$ is again the correlation function which behaves as 
$t^{-1/4}$ as the Lifshitz tricritical point is approached.
Again, this can be expressed in a scaling form
$$ \eqalignno{I(t,q)&=t s(q\xi)\cr
              s(x)&=(1+K^{\prime\prime}px^2+x^4)^{-1},&({\rm H.}27)\cr}$$
with $p$ fixed.    
Therefore, the presence of the Lifshitz tricritical behavior makes itself 
known by the fact that, in a fit of the structure function at long 
wavelengths, terms of order $q^4$ cannot be neglected with respect to terms of 
order $q^2$.
Again, when the effects of fluctuations are included, one expects the 
structure function to take a scaling form
$$I(q,T,b,g_0)=t^{\gamma}s\{q\xi(t,{b\over t^{1/\phi_t}},{g_0\over 
t^{1/\phi_g}}\}, \eqno({\rm H.}28)$$
but the form of the scaling function will differ, and the critical exponent 
$\gamma$ and the crossover exponents $\phi_t$ and $\phi_g$ will differ from 
their mean-field values of 1, 2, and 2. Experiments on mixtures of 
homopolymers have observed the change in the exponent $\gamma$ from its 
mean-field value of 1 to its
fluctuation driven value of 1.26 as the consolute line was approached
(section III).
Presumably, a similar change could be 
observed if the Lifshitz tricritical point were approached,
 because the region of 
fluctuation dominated behavior is somewhat larger.
Again, such an observation would be most interesting 
as there exist no calculated value for 
this exponent characterizing the Lifshitz tricritical point.

\centerline{\bf Appendix I: The ideal averages for the rigid-flexible
diblocks}

Below we show the results of the calculations of the ideal averages.
Here $a=\sin(ql)/ql$, f is the fraction of A monomers in the copolymer
and N=N$_A$+N$_B$ is the number of all the monomers in a copolymer.
We find,
$$\displaystyle{{1\over V}
<\hat\phi_A^{(n)}({\bf q})\hat\phi_A^{(n)}(-{\bf q})
>_0={1\over{\rho_0 N}}\left(
{{(1+a)}\over {(1-a)}}
fN-{2a\over {(1-a)^2}}(1-a^{fN})\right)}.\eqno({\rm I.}1)$$
$$\displaystyle{\eqalign{{1\over V}
<\hat\phi_A^{(n)}({\bf q})\hat\phi_B^{(n)}(-{\bf q})>_0=
{1\over {\rho_0 N}}{{1-a^{fN}}\over {1-a}}
&\int_0^1 dx \Biggr({{\sin (ql(1-f)Nx/2)}
\over {\sin (qlx/2)}}\cr &\times\cos (ql[(1-f)N+1]x/2)\Biggr)\cr}};\eqno({\rm I.}2)$$
$$\displaystyle{{1\over V}
<\hat\phi_B^{(n)}({\bf q})\hat\phi_B^{(n)}(-{\bf q})>_0=
{1\over {\rho_0 N}}\int_0^1 dx {{\sin^2 (ql(1-f)Nx/2)}\over {\sin^2 (qlx/2)}}};
\eqno({\rm I.}3)$$
$$\displaystyle{\eqalign{&{1\over V}
<\hat\phi_A^{(n)}({\bf q})\hat Q_{xx}^{(B)}
(-{\bf q})>_0=
{1\over V}
<\hat\phi_A^{(n)}({\bf q})\hat Q_{yy}^{(B)}(-{\bf q})>_0=\cr
&{1\over {\rho_0 N}}{{1-a^{fN}}\over {1-a}}\int_0^1 dx\Biggr(\Biggr
({1\over 4}-
{3\over 4}x^2\Biggr){{\sin (ql(1-f)Nx/2)}
\over {\sin (qlx/2)}}\cos (ql[(1-f)N+1]x/2)\Biggr)\cr}};\eqno({\rm I.}4)$$
$$\displaystyle{\eqalign{{1\over V}
<\hat\phi_B^{(n)}({\bf q})\hat Q_{xx}^{(B)}(-{\bf q})>_0=
&{1\over V}
<\hat\phi_B^{(n)}({\bf q})\hat Q_{xx}^{(B)}(-{\bf q})>_0=\cr
&{1\over {\rho_0 N}}\int_0^1 dx \left({1\over 4}-{3\over 4}x^2\right)
{{\sin^2 (ql(1-f)Nx/2)}\over {\sin^2 (qlx/2)}}\cr}};\eqno({\rm I.}5)$$
$$\displaystyle{\eqalign{{1\over V}
&<\hat Q_{xx}
^{(B)}({\bf q})\hat Q_{xx}^{(B)}(-{\bf q})>_0=
{1\over V}
<\hat Q_{yy}^{(B)}({\bf q})\hat Q_{yy}^{(B)}(-{\bf q})>_0=\cr
&{1\over {2\rho_0 N}}\int_0^1 dx \left
({9\over {16}}(1-x^2)^2-{3\over 2}(1-x^2)+
{1\over 2}\right)
{{\sin^2 (ql(1-f)Nx/2)}\over {\sin^2 (qlx/2)}}\cr}};\eqno({\rm I.}6)$$
$$\displaystyle{\eqalign{{1\over V}&<\hat Q_{xx}
^{(B)}({\bf q})\hat Q_{yy}^{(B)}(-{\bf q})>_0=\cr
&{1\over {2\rho_0 N}}\int_0^1 dx \left(
{27\over {16}}(1-x^2)^2-{3\over 2}(1-x^2)+
{1\over 2}\right)
{{\sin^2 (ql(1-f)Nx/2)}\over {\sin^2 (qlx/2)}}\cr}}.\eqno({\rm I.}7)$$
Finally in Eqs.(I.1-I.7) 
we take a limit of $q\rightarrow 0$, $N_A,N_B\rightarrow\infty$
such that $f=N_A/(N_A+N_B)=$const and $qlN=$const.  

\vfill\eject

\centerline{\bf References}

\item{$^{1)}$} P.G. de Gennes {\it Scaling Concept in Polymers Physics},
(Cornell University, Ithaca, N.Y. 1979) p 103, 114-115.
\item{$^{2)}$} P.G. de Gennes,{\it  J.Physique} {\bf 31}, 235 (1970);{\it Scaling 
Concept in Polymer Physics}
, (Cornell University Ithaca, N.Y.
1979) p 259.
\item{$^{3)}$} P.G. de Gennes,{\it J.Physique Lett.} {\bf 38}, 441 (1977).
\item{$^{4)}$} K.Binder, {\it Phys.Rev A} {\bf 29}, 341 (1984). 
\item{$^{5)}$} J.F.Joanny {\it J.Phys. A} {\bf 11}, L117 (1978).
\item{$^{6)}$} P.Pfeuty and G.Toulouse, {\it Introduction to the Renormalization
Group and to Critical Phenomena} (Wiley, New York, 1977) p 38. 
\item{$^{7)}$} D.J. Amit, {\it Field Theory, The Renormalization 
Group, and Critical Phenomena}, (World Scientific, Singapore 1984) p 87,109.
\item{$^{8)}$} Ch. Herkt-Maetzky and J. Schelten, {\it 
Phys.Rev.Lett} {\bf 51}, 896
(1983).
\item{$^{9)}$} D.Schwahn, K. Mortensen and H. Yee-Madeira, {\it Phys.Rev.Lett.}
\hfill\break {\bf 58}, 1544 (1987). 
\item{$^{10)}$} F.S. Bates, J.H. Rosendale, P. Stepanek, T.P. Lodge,
P.Wiltzius, G.H. Fredrickson and R.P. Hjelm Jr, {\it Phys.Rev.Lett.} {\bf 65},
1893 (1990).
\item{$^{11)}$}G.Meier, B.Momper and E.W.Fischer,{\it J.Chem.Phys.} {\bf 97},
5884, (1992).
\item{$^{12)}$} M.D. Gehlsen, J.H.Rosendale, F.S.Bates, G.D.Wignall,
L. Hansen  and K. Almdal, {\it Phys.Rev.Lett.} {\bf 68}, 2452, (1992).
\item{$^{13)}$} G. Meier, D.Schwahn, K. Mortensen and S. Janssen,
{\it Europhys. Lett.} {\bf 22}, 577 (1993).
\item{$^{14)}$} R.Ho\l yst and T.A.Vilgis, 
{\it J.Chem.Phys.} {\bf 99}, 4835 (1993).
\item{$^{15)}$} R.Ho\l yst, {\it Phys.Rev.Lett.} {\bf 72}, 2304 (1994).
\item{$^{16)}$} M.G.Brereton and T.A.Vilgis, {\it J.Phys. (France)},
{\bf 50}, 245 (1989).
\item{$^{17)}$} R.Ho\l yst and T.A.Vilgis, {\it Phys.Rev. E} {\bf 50}, 2087
 (1994).
\item{$^{18)}$} T.A.Vilgis and G.Meier, {\it
J.Phys. I (France)}, {\bf 4}, 985 (1994).
\item{$^{19)}$} A. Sariban and K.Binder, {\it 
Macromolecules} {\bf 21}, 711 (1988).
\item{$^{20)}$} J.L.~Barrat and G.~H.~Fredrickson, {\it J.Chem.Phys.} {\bf 95},
1281 (1991). 
\item{$^{21)}$} H.Fried and K.Binder, {\it J.Chem.Phys.} {\bf 94}, 8349 (1991);
A.Weyersberg and T.A.Vilgis, {\it Phys.Rev.E} {\bf 48}, 377 (1993);
A.Gauger, A.Weyersberg and T.Pakula, {\it Macromol. Chem.Theory Simul.}
{\bf 2}, 531 (1993).
\item{$^{22)}$} L.Leibler, {\it Macromolecules}, {\bf 13}, 1602 (1980).
\item{$^{23)}$} G.H.Fredrickson an E.Helfand, {\it J.Chem.Phys.} {\bf 87},
697 (1987).
\item{$^{24)}$} M.W.Matsen and M.Schick, {\it Phys.Rev.Lett.} {\bf 72}, 2660
(1994).
\item{$^{25)}$} M.W.Matsen and M.Schick, {\it Macromolecules} {\bf 27}, 4014
(1994).
\item{$^{26)}$} E.Helfand {\it J.Chem.Phys.} {\bf 62}, 999 (1975).
\item{$^{27)}$} K.M.Hong and J.Noolandi {\it Macromolecules}, {\bf 14}, 727
(1981).
\item{$^{28)}$} F.S.Bates and G.H.Fredrickson, {\it Annu.Rev.Phys.Chem.}
{\bf 41}, 525 (1990);\hfill\break F.S.Bates {\it Science} {\bf 251}, 898 (1991). 
\item{$^{29)}$} G.E.Molau {\it Block Copolymers} (ed. S.L. Aggarawal)
(Plenum Press, New York ) (1970). 
\item{$^{30)}$} E.L.Thomas et al, {\it Macromolecules} {\bf 19}, 2197 (1986);
D.S.Herman et al {\it Macromolecules} {\bf 20}, 2940 (1987).
\item{$^{31)}$} H.Hasegawa et al, {\it Macromolecules} {\bf 20}, 1651 (1987).
\item{$^{32)}$} E.L.Thomas et al , {\it Nature} {\bf 334}, 598 (1988).
\item{$^{33)}$} D.M. Anderson, H.T. Davis,J.C.C. Nitsche, L.E.Scriven,
{\it Adv.Chem.Phys.} {\bf 77}, 337 (1990).
\item{$^{34)}$} D.Broseta and G.H.Fredrickson, {\it J.Chem.Phys.}
{\bf 93}, 2927 (1990).
\item{$^{35)}$} R.Ho\l yst and M.Schick, {\it J.Chem.Phys.} {\bf 96}, 7728 (1992).
\item{$^{36)}$} M.W.Matsen and M.Schick, {\it Macromolecules} 
{\bf 26}, 3878 (1993).
\item{$^{37)}$} M.W.Matsen, {\it Phys.Rev.Lett.} 
{\bf 74}, 4225 (1994).
\item{$^{38)}$} G.Gompper and M.Schick, 
{\it Self Assembling Amphiphilic Systems},
vol. {\bf 16} {\it Phase Transitions and Critical Phenomena} eds. C.Domb and
J.L.Lebowitz, Academic Press (1994).
\item{$^{39)}$} J.~Stephenson {\it J.~Math.~Phys.} {\bf 11}, 420 (1970); M.~E.~Fisher and
B.~Widom,\hfill\break {\it  J.~Chem.~Phys.} {\bf 50}, 3756 (1969).
\item{$^{40)}$} G.H.Fredrickson and S.T.Milner, {\it Phys.Rev.Lett.}
{\bf 67}, 835 (1991).
\item{$^{41)}$} G.Grinstein in {\it Fundamental Problems in Statistical
Mechanics IV} ed E.G.D. Cohen (North Holland, Amsterdam) (1985).
\item{$^{42)}$} S.F.Edwards and P.W.Anderson, {\it J.Phys. F} {\bf 5}, 965 (1975).
\item{$^{43)}$} M.Mezard, G.Parisi and M.A. Virasoro, {\it Spin Glass Theory
and Beyond}, (World Scientific, Teaneck, NJ) (1987).
\item{$^{44)}$} M.W.Matsen, {\it J.Chem.Phys.} {\bf 102}, 3884 (1995).
\item{$^{45)}$} M.D.Gehlsen , K.Almdal and F.S.Bates, {\it Macromolecules}
{\bf 25}, 939 (1992). 
\item{$^{46a)}$} M.E.Cates and J.M.Deutsch, {\it J.Physique} {\bf 47},
2121 (1986). 
\item{$^{46b)}$} M.G.Brereton and T.A.Vilgis, {\it J.Phys.A} {\bf 28},
1149 (1995).
\item{$^{47)}$} A.Weyersberg and T.A.Vilgis, {\it Phys.Rev.E} {\bf 49}, 3097
(1994).
\item{$^{48)}$} J.F.Marko, {\it Macromolecules} {\bf 26}, 1442 (1993).
\item{$^{49)}$} M.Benmouna, R.Borsali and H.Benoit, {\it J.Phys. II}
(France) {\bf 3}, 1041 (1993). 
\item{$^{50)}$} R.Ho\l yst and M.Schick, {\it J.Chem.Phys.} {\bf 96}, 730 (1992).
\item{$^{51)}$} P.G. de Gennes and J.Prost, 
{\it The Physics of Liquid Crystals}, (Clarendon Press, Oxford)
p57 (1993).   
\item{$^{52)}$} C.Singh, M.Goulian, A.J.Liu and G.H.Fredrickson,
{\it Macromolecules} {\bf 27}, 2974 (1994).
\item{$^{53)}$} {\it Recent Advances in Liquid Crystalline Polymers}, edited by L. 
Chapoy, (Elsevier, New York 1985).
\item{$^{54)}$} {\it Polymer Liquid Crystals}, edited by A. Ciferri, W.R. Kingbaum, 
and R.B. Meyer (Academic, New York 1982);
{\it Polymeric Liquid Crystals}, edited by A. Blumstein, (Plenum 
Press, New York, 1985).
\item{$^{55)}$} E. Bianchi, A. Ciferri, and A. Tealdi,{\it  Macromolecules} {\bf 15}, 1268 
(1982);\hfill\break C. Casagrande, M. Veyssi\"e, and H. Finkelmann, {\it J. Phys. Lett.} 
{\bf 43}, L671 (1982); F. Auriemma, P. Corradini, A. Roviello, and M. 
Vacatello, {\it Eur. Polym. J.} {\bf 1}, 57 (1989).
\item{$^{56)}$} J.S. Moore and S.I. Stupp, {\it Macromolecules} 
{\bf 21}, 1217 (1988).
\item{$^{57)}$} P. Ma\"issa and P. Sixou, {\it Liquid Crystals} 
{\bf 5}, 1861 (1989).
\item{$^{58)}$} N. Sait\^o, K. Takahashi, and Y. Yunoki, 
{\it J. Phys. Soc. Jap.} {\bf 
22}, 219 (1967). 
\item{$^{59)}$} G.H. Fredrickson and L. Leibler, {\it Macromolecules} 
{\bf 23}, 531 
(1990).
\item{$^{60)}$} R.Ho\l yst and M.Schick, {\it J.Chem.Phys.} {\bf 96},
721 (1992).      
\item{$^{61)}$} M.Lifschitz, J.Dudowicz and K.F.Freed, {\it J.Chem.Phys}
{\bf 100}, 3957 (1994).  
\item{$^{62)}$} M.Benmouna, T.A.Vilgis, M.Daoud and M.Benhamou,
{\it Macromolecules} {\bf 27}, 1172 (1994).
\item{$^{63)}$} M.G.Brereton and T.A.Vilgis, {\it Macromolecules}  {\bf 23}, 
2044 (1990).
\item{$^{64)}$} R.Borsali and T.A.Vilgis, {\it J.Chem.Phys.} {\bf 93}, 3610 
(1990).
\item{$^{65)}$} U.Genz and T.A.Vilgis, {\it J.Chem.Phys.} {\bf 101}, 7101 
(1994).  
\item{$^{66)}$} T.A.Vilgis and U.Genz, {\it J. Phys I (France)} {\bf 4}, 1411 
(1994). 
\item{$^{67)}$} G.H.Fredrickson and K.Binder, {\it J.Chem.Phys.} {\bf 91},
7265 (1989).
\item{$^{68)}$} U.Genz and T.A.Vilgis, {\it J.Chem.Phys.} {\bf 101}, 7111 
(1994).  
\item{$^{69)}$} K. Freed, {\it Renormalization Group Theory of Macromolecules}, 
(Wiley, New York 1987), p21. 
\item{$^{70)}$} S.F. Edwards, {\it 
Proc. Phys. Soc. (London)} {\bf 88}, 265 (1966).
\item{$^{71)}$} K.S.Schweitzer and J.G.Curro, {\it Phys.Rev.Lett.} {\bf 60},
809 (1988).
\item{$^{72)}$} J.-P. Hansen and I.R.MacDonald, {\it Theory of Simple
Liquids}, 2nd edition (Academic Press) (1986).
\item{$^{73)}$} R.Evans, {\it Adv. Phys.} {\bf 28}, 143 (1979).
\item{$^{74)}$} F.S.Bates, P.Wiltzius and G.H.Fredrickson,
{\it Phys.Rev.Lett.} {\bf 72}, 2305 (1994).
\item{$^{75)}$} M.Doi and S.F.Edwards, {\it The Theory of Polymer Dynamics},
(Clarendon Press, Oxford, 1986) p. 22,23. 
\item{$^{76)}$} T.Ohta and K.Kawasaki, {\it Macromolecules} {\bf 19}, 2621
(1986).  
\item{$^{77)}$} S.Brazovskii, {\it Sov.Phys.JETP} {\bf 41}, 85 (1975).
\item{$^{78)}$} (a) K.Almdal, J.H.Rosendale, F.S.Bates, G.D.Wignall and G.H.
Fredrickson, \hfill\break {\it Phys.Rev.Lett.} {\bf 65}, 1112 (1990);
(b) V.T.Bartels, M.Stamm, V.Abetz and K.Mortensen; {\it Europhys.Lett.}
{\bf 31}, 81 (1995).
\item{$^{79)}$} D.A.Hajduk et al, {\it Macromoecules}, {\bf 27}, 4063
(1994); M.Schulz et al, \hfill\break {\it Phys.Rev.Lett.} {\bf 73}, 86 (1994).
\item{$^{80)}$} H.Benoit and G.Hadziioannou, {\it Macromolecules}
{\bf 21}, 1449 (1988). 
\item{$^{81)}$} T.Pakula and S.Geyler, {\it Macromolecules}, {\bf 21},
1665 (1988).
\item{$^{82)}$} S.Geyler and T.Pakula, {\it Macromol. Chem. Rap. Commun.},
{\bf 9},
531 (1988).
\item{$^{83)}$} K.Mori, H.Tanaka and T.Hashimoto, {\it Macromolecules}
{\bf 20}, 381 (1987).
\item{$^{84)}$} A.M.Mayes and M.J. Olvera de la Cruz, {\it J.Chem.Phys.}
{\bf 91}, 7228 (1989).
\item{$^{85)}$} A.M.Mayes and M.J. Olvera de la Cruz, {\it J.Chem.Phys.}
{\bf 95}, 4670 (1991).
\item{$^{86)}$} M.W.Matsen and M.Schick, {\it Macromolecules}
{\bf 27}, 187 (1994).
\item{$^{87)}$} G.Hadziioannou and A.Skoulios, {\it Macromolecules}
{\bf 15}, 258 (1982).
\item{$^{88)}$} M.D.Gehlsen, K.Almdal and F.S.Bates, {\it Macromolecules}
{\bf 25}, 939 (1992).
\item{$^{89)}$} R.W.Richards and J.L.Thomason, {\it Macromolecules}
{\bf 16}, 16 (1983).
\item{$^{90)}$} E.I.Shakhnovich and A.M.Gutin, {\it J.Phys. (Paris)}
{\bf 50}, 1843 (1989).
\item{$^{91)}$} {\it Physics of Amphiphilic Layers}, ed. J.Meunier,
D.Langevin and N.Boccara (Springer, Berlin ) (1987).
\item{$^{92)}$} L.Leibler, {\it Macromolecules}, {\bf 15}, 1283 (1982).
\item{$^{93)}$} G.Gompper and M.Schick, {\it Phys.Rev.B} {\bf 41}, 9148
(1990). 
\item{$^{94)}$} M.Mueller and M.Schick (unpublished).
\item{$^{95)}$} M.P.Allen, G.T.Evans, D.Frenkel and B.M.Mulder,
{\it Adv. Chem.Phys.} \hfill\break{\bf XXXVI}, 
ed. I.Prigogine and S.A. Rice
(John Wiley \& Sons Inc 1-166 (1993).
\item{$^{96)}$} A.Poniewierski and R.Ho\l yst, {\it Phys.Rev.Lett.}
{\bf 61}, 2461 (1988); {\it Phys.Rev. A} {\bf 41}, 6871 (1990).
\item{$^{97)}$} L.Onsager {\it Proc. NY Acad. Sci.} {\bf 51}, 627 (1949).
\item{$^{98)}$} F.Brochard, J.Jouffroy and P.Levinson, {\it J.Phys.}
{\bf 45}, 1125 (1984).
\item{$^{99)}$} C.Shen and T.Kyu, {\it J.Chem.Phys. } {\bf 102}, 556 (1995).
\item{$^{100)}$} H.Chiu and T.Kyu, {\it Macromolecules} (in press) (1995).
\item{$^{101)}$} 
M. Warner and P.J. Flory, {\it J. Chem. Phys.} {\bf 73}, 6327 (1980).

\vfill\eject

\centerline{\bf Figure Captions}

\item{Fig.1} The schematic picture of the 
unit cell of cubic bicontinuous double diamond phase formed in AB diblock
copolymers. The 
junction of the A-B blocks are located at the surface shown in the figure.
The surface defines the boundary between  A rich and B rich domains.
Symmetry Pn$\bar 3$m. The experimental studies suggest that the
surface describing the boundary of domains in diblock copolymers 
is of constant mean curvature and is determined by the area minimization 
subject to fixed volume fraction of A and B blocks.
\item{Fig.2} The schematic picture  of the 
unit cell of gyroid structure formed in AB diblock
copolymer systems. The 
junction points of the A-B blocks are located at the  surface shown in the
figure.
Symmetry Ia$\bar 3$d.   
\item{Fig.3} Phase diagram of the polystyrene-polyisoprene system$^{30,31)}$.
The bcc phase is the body center cubic phase, where droplets of the minority component
in the diblock copolymer
form an ordered lattice. The  hex phase 
is the hexagonal phase where the  parallel
cylinders of the minority component surrounded by the majority component form
a hexagonal lattice. The obdd phase is the ordered bicontinuous double diamond
structure consisting of two channels each of diamond symmetry
separated by the surface (Fig.1). ODT stands for order-disorder transition.
$\chi$ is the Flory interaction parameter, $N$ is the polimerization index and
$f$ is the fraction of one component in the diblock copolymer system. 
The region of stability of the  
gyroid phase (Fig.2) which has been recently observed in this system
is not shown here. 
After Ref. 28.   
\item{Fig.4} Bridge and loop configurations in the lamellar phase of multiblock 
copolymer system. After Ref. 44.
\item{Fig.5} {\bf (a)} On the left is shown the permitted configuration of an 
unknotted ring. It cannot turn into the knotted configuration shown
on the right (not vice versa). {\bf (b)} Two unknotted rings are shown in the
allowed configuration.
It cannot turn into the knotted configuration shown on the right. After Ref.46. 
\item{Fig.6} Microscopic model of an AB diblock copolymer molecule with
different stiffness of the blocks. The size of the A segment, $b_A$,
is much smaller than the size of the B segment $b_B$.After Ref.52.    
\item{Fig.7} The one-loop diagrams contributing to the
two-body vertex function. The solid lines represent $1/\Gamma_2^{(0)}$
while the higher order vertex functions 
are represented
by points in the diagrams. The n-body vertex function has n lines 
emanating from the point representing this function.
The dashed lines
are not parts of the diagrams
but are included only for clarity (according to Ref.(7)).
Apart from $1/\Gamma_2^{(0)}$, 
only $\Gamma_3^{(0)}$ and $\Gamma_4^{(0)}$ appear
in the structure of the diagrams. 
\item{Fig.8} The one-loop diagrams contributing to the three-body
vertex function. The legend as in Fig. 1. Apart from $1/\Gamma_2^{(0)}$
only
$\Gamma_3^{(0)}$ and $\Gamma_5^{(0)}$ appear in the structure of the
diagrams.    
\item{Fig.9} The plot of the logarithm of the scattering intensity versus
the logarithm of the temperature difference. After Ref.11.   
Close to the critical point the intensity diverges according to the power law
$\vert T-T_c\vert^{-\gamma}$. The experiment was performed here on the symmetric
mixture of the PDMS and PEMS.   
\item{Fig.10} The inverse of the scattering intensity at zero q vector.
The temperature $T_x$ marks the Ginzburg region where there crossover
from the mean-field to the non-mean-field behavior. $T_b$ is the
critical temperature on this plot. Please note that $T_{s,mf}$ 
(mean-field temperature) is {\bf not correctly} identified
(see text and Ref.15,74). After Ref.10.  The experimental system studied in 
this paper is PI-PEP mixture.      
\item{Fig.11} The scattering intensity$^{22})$ for the diblock copolymer
in the random phase approximation for the 
system with composition $f=0.25$ as a function of $q^2R^2$, where
$R^2=Nl^2/6$ is the radius of gyration; solid line ($\chi N=12.5$),
dashed line ($\chi N=16.0$) and the dotted line ($\chi N=17.5$).
The peak in the scattering intensity is fixed and does not change with
temperature, which is here proportional to $\sim 1/\chi$.
\item{Fig.12} The phase diagram for the diblock copolymer system in the
random phase approximation (after Ref.22 and Ref.23). 
The diagram is symmetric with respect to f=0.5 (symmetric diblock copolymer).
The transition to the body centered cubic (BCC) and hexagonal
(HEX) structure is first order. The direct second order
transition from the disordered phase (DIS) to the lamellar phase occurs 
only at $f=0.5$. 
\item{Fig.13} The phase diagram for the diblock copolymer system in the
self consistent one-loop approximation (Hartree) (after Ref.23) for
$N=10^6$. 
Compare
with Fig.12; in the limit $N\to\infty$ the results of the
Hartree approximation reduce to the random phase approximation. 
The dashed line marks the divergence of the scattering intensity
in the random phase approximation (of course in the one-loop approximation
for finite $N$ the structure factor does not diverge at this line). 
\item{Fig.14} Peak position, $q^*$, (in \AA$^{-1}$)
versus the polimerization index, $N$ for the PEP-PEE diblock copolymer
system at fixed temperature T=296 K. After Ref.20. The vertical solid
line indicates the microphase separation obtained in the theory.
The solid line of slope -1/2 indicates the result of the random 
phase approximation. Open squares are the predictions of the one-loop
approximation theory and full triangles are the experimental results$^{78a)}$.
\item{Fig.15} The phase diagram obtained in the self consistent field
theory. Here C is the cubic bcc phase, H is the hexagonal phase, L is the
lamellar phase and G is the gyroid phase (Fig.2). 
The diagram is symmetric with
respect to reflection at f=0.5. After Ref.24. The transition 
from the disordered to the
lamellar phase is second order; exactly as in the random phase
approximation. It is marked by a dot in the phase diagram. All the
other transitions are first order.
\item{Fig.16} The phase diagram of the diblock copolymer system
with conformational asymmetry. The size of the A monomer, $l_A$, is
equal to $\sqrt{10}l_B$, where $l_B$ is the size of the B monomer.
The legend as in Fig.15, where $l_A=l_B$. Please note that
conformational asymmetry does not change the topology of the
phase diagram in this theory. After Ref[25].
\item{Fig.17} The scattering intensity,
calculated in the random phase approximation,
for the symmetric ring diblocks and
linear diblock copolymer system at infinite temperature $\chi =0$.
The scattering from rings is less intense and the peak is at higher
q vectors. After Ref.48.
\item{Fig.18} The instability towards the ordered phase (divergence
of the scattering intensity). The transition temperature ($\sim 1/\chi$)
is approximately 40\% lower in the rings than in the linear diblocks.
After Ref.48.
\item{Fig.19} Comparison of the radii of gyration for linear and
ring (cyclic) symmetric diblock copolymers of length $N=20$.
After Ref.47.
Here $T_c$ is the temperature of the microphase separation 
(to the lamellar phase); $\epsilon$ is the enegy interaction parameter
between the monomers and $a$ is the monomer size.
The inverse of this temperature, in the
random phase approximation, is shown (Fig.18) for $f=1/2$.
\item{Fig.20} The distance between the centers of masses of A and B 
blocks in the ring diblock copolymer. After Ref.47.
It indicates that the rings strongly stretch as the temperature
is lowered towards the temperature of the microphase separation, $T_c$.
\item{Fig.21} The radius of gyration for the single block in
the ring copolymer $\alpha =A,B$. The blocks shrink as we lower the
temperature , similarly as in the case of linear diblocks. 
$T_c$ is the temperature of the microphase separation. After Ref.47. 
\item{Fig.22} Phase diagram for the ABA triblock copolymer system
calculated in the random phase approximation. Here BCC is the 
body center cubic phase, HCP is the hexagonal phase and LAM is the
lamellar phase. After Ref.84.
The two A blocks have the same number of A monomers.
\item{Fig.23} Phase diagram of the tenary mixture at equal A and B 
homopolymer concentrations. 
Here only instability lines are marked. The three phase coexistance 
region is marked only schematically.
In the region named {\it lamellar} there
should be more different ordered phases, but they have not been studied,
here. Also note that the point where all lines meet is the Lifshitz
tricritical point. After Ref.35. See also Ref.34 and 36.
\item{Fig.24} The same phase diagram as shown in Fig.23. Here the
lines in the disordered phase shows the various stages of 
local ordering. LD denotes the disorder line, LF is the Lifshitz line 
for the $\Psi\Psi$ correlation function and
LE and EG are the Lifshitz and  the equimaxima lines
of the $\eta\eta$ structure function. 
L is the Lifshitz tricritical point.
The region between the LD and LF shows the region where the
correlation function starts to oscillate showing a local ordering.
After line LF the ordering becomes dominant and the peak in the
structure functions appear. The region between the LG line and LE-EG
is the region where the homopolymers order and where their structure
functions develop a peak. At the EG line the homopolymer homopoymer
structure function shows two peaks et equal height, one at zero
q vector and one at non zero q vector. After Ref[35].
\item{Fig.25} The stability limits of the disordered isotropic phase
(I) against lamellar (L) (solid line) 
or nematic (N) perturbations (dashed line)
in the rigid-flexible diblock copolymer system for (a) $v_{BB}/\chi =0$,
(b)  $v_{BB}/\chi =0.4$, (c)  $v_{BB}/\chi =3.0$. Dot denotes the
tricritical point. After Ref.50.
\item{Fig.26} The structure function $\tilde G_{\Psi\Psi}(q_z)$ 
(case ({\bf (a),(c)}) and its one dimensional Fourier transform 
(density-density correlation function) 
$G_{\Psi\Psi}(z)$ (case ({\bf (b),(d)}). Here N is the number of monomers
in a diblock copolymer, $\rho_0$ is the monomer density and $N_Bl=
(1-f)Nl$ is the length of the rigid part.
{\bf (a),(b)}f=0.2, $\chi N=$19.51, and
$v_{BB}/\chi=0$ (dashed line), and f=0.2, $\chi N=$19.51,
$v_{BB}/\chi=$0.4 (solid line). 
{\bf (c),(d)} f=0.62, $v_{BB}/\chi=0$,
$\chi N$=11.75 (dashed line) and f=0.62, $v_{BB}/\chi$=3.0,
$\chi N$=11.35 (solid line).
The parameters for the solid line are
close to those of the tricritical point shown in Fig.25. After Ref.50.
\item{Fig.27} The structure function $\tilde G_{QQ}(q_z)$ versus $q_z$
and its one dimensional Fourier transform (nematic order parameter-nematic 
order parameter correlation function) $G_{QQ}(z)$ versus z for 
f=0.62. The parameters for the solid and dashed lines are as in 
Fig(26c,d). After Ref.50.
\item{Fig.28} The structure function $\tilde G_{\Psi Q}(q_z)$ 
versus $q_z$ and its
one dimensional Fourier transform (density-nematic order parameter correlation
function) $G_{\Psi Q}(z)$ versus z for the same two systems as in 
Fig.26; {\bf (a),(b)} f=0.2; {\bf (c),(d)} f=0.62. After Ref.50.

\vfill\eject\end